\documentclass[useAMS,usenatbib]{mn2e}
\usepackage{graphics,graphicx,epsfig,psfig}
\usepackage[normalem]{ulem}
\usepackage[dvipsnames]{color}
\usepackage{soul,xcolor}
\usepackage{amsmath, amssymb}
\usepackage[]{inputenc,amssymb}
\usepackage{booktabs,caption}
\usepackage{amsmath}

\setstcolor{red}

\def \be{\begin{equation}}
\def \ee{\end{equation}}
\def \ba{\begin{eqnarray}}
\def \ea{\end{eqnarray}}
\def \etal{{et al.}}

\definecolor{webgreen}{rgb}{0,.5,0}
\definecolor{webbrown}{rgb}{.6,0,0}
\usepackage[%
    colorlinks = true,%
    linkcolor = blue,%
    urlcolor  = blue,%
    citecolor = webgreen,%
    anchorcolor = blue]{hyperref}

\newcommand{\ufhref}[3][blue]{\href{#2}{\color{#1}{#3}}}%

\def \kmps{km s$^{-1}$}

\def \etal{\textit{et. al. }}

\definecolor{webgreen}{rgb}{0,.5,0}
\definecolor{webbrown}{rgb}{.6,0,0}

\usepackage[T1]{fontenc}
\usepackage{ae,aecompl}

\usepackage{newtxtext,newtxmath}
\title[Extra-planar X-ray in spiral galaxies]{Extra-planar X-ray emission from disc-wide outflows in spiral galaxies}

\author[Vijayan \etal]
{
\parbox{\textwidth}{Aditi Vijayan$^{1,2}$\thanks{Email: aditiv@rri.res.in},
 Kartick C. Sarkar$^{1,2,4}$, 
 Biman B. Nath$^1$, 
 Prateek Sharma$^2$ and
 Yuri Shchekinov$^{1,3}$} \vspace{0.4cm}\\
\parbox{\textwidth}{ 
$^{1}$Raman Research Institute, Bangalore, 500080, India\\
$^{2}$Joint Astronomy Programme and Department of Physics, Indian Institute of Science, Bangalore 560012, India\\
$^3$ Lebedev Physical Institute of Russian Academy of Sciences, ASC, Moscow 117997, Russia\\
$^4$ Racah Institute of Physics, Hebrew University of Jerusalem, Jerusalem, Israel}
}
\date{Last updated 2015 May 22; in original form 2013 September 5}

\pubyear{2015}

\begin{document}
\label{firstpage}
\pagerange{\pageref{firstpage}--\pageref{lastpage}}
\maketitle

\begin{abstract}
We study the effects of mass and energy injection due to OB associations spread across the rotating disc of a Milky Way-type 
galaxy, with the help of 3D hydrodynamic simulations. We compare the resulting X-ray emission with that produced from the injection of 
mass and energy from a central region. We
find that the predicted X-ray image shows a filamentary structure that arises even in the absence of disc gas inhomogeneity.
This structure stems from warm clumps made of disc material being lifted by the injected gas. 
We show that as much as half  of the total X-ray emission comes from regions surrounding warm clumps that are made of a mix 
of disk and injected gas. This scenario has the potential to explain the origin of the observed extra-planar X-ray emission around star forming galaxies and can be used to understand the observed sublinear relation between the $L_X$, the total X-ray luminosity, and SFR.
We quantify the mass contained in these `bow-shock' regions. We also show that the top-most region of the outer shock above the central area emits harder X-rays than the rest. Further, we find that the mass distribution in different temperature ranges is bimodal, peaking at $10^4\hbox{--}10^5$ K (in warm clumps) and $10^6\hbox{--}10^7$ K (X-ray emitting gas). The mass loading factor
is found to decrease with increasing SFR, consistent with previous theoretical estimates and simulations.
\end{abstract}


%%%%%%%%%%%%%%%%%%%%%%%%%%%%%%%%%%%%%%%%%%%%%%%%%%

%%%%%%%%%%%%%%%%% BODY OF PAPER %%%%%%%%%%%%%%%%%%

% The MNRAS class isn't designed to include a table of contents, but for this document one is useful.
% I therefore have to do some kludging to make it work without masses of blank space.
%\begingroup
%\let\clearpage\relax
%\tableofcontents
%\endgroup
%\newpage

\section{Introduction}
Outflows from galaxies are believed to regulate star formation in galaxies by ejecting gas and therefore influence 
galactic evolution (e.g. \cite{Dekel1986}). Moreover, these outflows can also determine the physical state of the
circumgalactic medium (CGM) \citep{Oppenheimer2006} and enrich the intergalactic medium (IGM) with metals (e.g., 
\cite{nath1997, ferrara2000, aguirre2001}).
Although there are many physical processes that can launch these outflows,
it is believed that thermal pressure arising from supernovae driven energy and mass deposition is the most important
process in star forming galaxies (e.g., \cite{larson1974}). In particular, large superbubbles generated by stellar winds and
supernovae in OB associations, where most massive stars are born, have been shown to be efficient in breaking through
the disc gas with enough impetus to reach large distances from the galaxy (e.g, \cite{nath2013,Sharma2014}).

X-ray observations of outflows have been useful in the interpretation of the galactic outflows phenomenon \citep{strickland2002, strickland2007}. This is because a significant fraction of the total mass and energy in the outflowing gas resides in the form of hot gas ($\ge 10^6$ K). In more recent observations by \cite{Li2013} of highly inclined ($i \gtrsim 60^\circ$) galaxies, a mysterious nature of the extra-planar soft X-ray emission is revealed. They notice that the soft X-ray luminosity ($L_X$) has only weak dependence on the stellar mass of the galaxy ($M_\ast$) compared to elliptical galaxies, indicating that the X-ray emitting material is not a gravitationally collapsed atmosphere. They also find that the temperature of the X-ray emitting plasma ($T_X$) is uncorrelated with the star formation rate which can further (loosely) mean that the emitting material is not shocked. The only remaining possibility is then the X-ray emission by the wind driven by supernovae activity at the galaxy.

The literature related to finding X-ray emission from galacitc wind goes back to \cite{chevalier1985} who estimated the X-ray luminosity of the nearby starburst galaxy M82 assuming that outflows are launched from a central region ($\sim 200$ pc). This scenario has the appeal of analytical tractability and simplicity, and has therefore been extensively used (e.g., \cite{Zhang2014, Thompson2016}). In this scenario, most of the X-ray emission arises from the central, hot and dense region (whose temperature is estimated by assuming that a fraction $\alpha$ of the total energy injection is thermalised).
%\sout{, and a shock-heated shell comprising of halo gas and injected gas (separated by a contact discontinuity)}. 

This scenario has also been used to constrain the wind parameters, like $\alpha$ and mass loading factor ($\beta$), using an observed linear relation between the total soft X-ray luminosity ($L_X$) in the disc and SFR \citep{Thompson2016, Bustard2016}. However, it faces serious limitations while explaining a sub-linear relation between the observed extra-planar X-ray luminosity and the SFR \citep{Strickland2004, Grimes2005, Wang2016} in highly inclined disc galaxies. In such galaxies, the soft X-ray emission from the central region of the wind is expected to be absorbed due to high column density along the line of sight. Moreover, the surface brightness of the extra-planar emission, in the wind model, falls off too rapidly as compared to recent observations by \cite{Li2013}.

Analytical wind models further assume that the wind from star forming region expands as a single fluid. In other words, the loaded mass is assumed to be mixed up with the wind instantaneously which, thereby, affects the thermal and dynamical features of the wind. Observations of nearby star forming galaxies, however, find a coexistence of H$\alpha$ emitting warm gas and X-ray emitting hot gas \citep{Grimes2005}. The thermodynamics of such a multiphase wind is difficult to obtain in analytical wind models and requires dedicated numerical simulations.

Another shortcoming of the typical wind models is that the star formation is assumed to be centralised ($\sim 200$ pc). The star formation in a disc galaxy, however, extends to regions beyond the central few hundred parsecs. According to the
well-known Kennicutt-Schmidt law, the surface density of star formation rate (SFR) is given by $\Sigma_{\rm SFR} \propto 
\Sigma_{\rm gas} ^{1.4}$, where $\Sigma_{\rm gas}$ is the surface density of gas \citep{kennicutt1998} (although the scaling
may vary from $\Sigma_{\rm SFR}\propto \Sigma_{\rm gas}$ at high surface densities to 
$\Sigma_{\rm SFR}\propto \Sigma_{\rm gas}^2$ at low surface densities \citep{bigiel2008}).
Therefore it is important to study the effects of distributed star formation in a 
disc galaxy on the morphology, dynamics, and emission characteristics in various bands of the outflowing gas. It then
becomes necessary to use 3-D hydrodynamical simulations for such a study. Previously, \cite{Cooper2008}%, Cooper2009} 
 studied the morphology and emission properties (in X-ray and optical bands) for outflows arising from a nuclear star forming disc with inhomogeneous 
distribution of interstellar medium (ISM), they concluded that the observed filamentary structure of outflows, as seen in
optical and X-ray images, is caused by the inhomogeneities in the ISM. 

However, the star forming region in their simulation
was confined to the central 150 pc, for a galaxy with disc mass $\sim 6 \times 10^9$ M$_\odot$ (and halo temperature
of $5$ million K). Interestingly they found that the bulk of X-rays are emitted from the gas in the filamentary structure of the
outflows, mostly in the bow-shocks created by dense clumps of disc gas lifted by the ram pressure of the hot wind. This is
inherently different from the scenario of a centralised injection of energy and mass, as explained above, although \cite{Cooper2008} 
admittedly studied only the X-ray emission close ($\le 1$ kpc) to the plane of the disc. 

%\sout{Recently, }\cite{Fielding2017} \sout{have used 3-D simulations to study the dynamics and morphology of outflows, arising from
%injection points distributed in the disc of a galaxy. In their simulation, the star formation regions were identified by using
%the Kennicutt-Schmidt law, and they studied the long-term effects (over more than 100 Myr) of such a dynamical linking of density and star %formation process, such as the relation between the mass loading factor and SFR and so on. }

In this paper, we address the issue of X-ray emission properties (as in \cite{Cooper2008}) from a distributed star formation scenario throughout the disc (unlike the case of central injection studied in \cite{Cooper2008}). We use a set of idealized 3-D hydrodynamical simulations, in which the star formation sites are chosen such as to make the surface density of SFR follow the Kennicutt-Schmidt law. 
Our main aim in this paper is to study the X-ray properties of outflows out to a few kpc from the plane of the galaxy. In light of the computationally expensive nature of such simulations (in particular, owing to the fact that our spatial resolution in the inner disc is of order $\sim 10$ pc, and our box size extends to $6$ kpc), we study the outflows for 10 Myr, during which time, for typical starburst parameters, the shell of swept out matter reaches a distance of a few kpc. We also do not dynamically link the star formation sites with the current (and evolving) density structure in the disc, and allow the star formation sites to fire up simultaneously and continue for 10 Myr.
Our set-up therefore complements the previous studies of \cite{Cooper2008} and \cite{Fielding2017}.

\begin{table}
\caption{Parameters used to set up a steady state gaseous distribution. The references for these parameters can be found in \citep{KCSI}.}
\centering
\begin{tabular}{ || p{3cm} | p{3cm} || } 
 \hline
 Parameters (units) & Value  \\ 
 \hline
 $M_{\rm vir}$ ($M_{\odot}$) & $10^{12}$ \\
 %\hline
 $M_{\rm disc}$($M_{\odot}$) & $5 \times 10^{10}$ \\
 %\hline
 $T_{\rm vir}$ (K) & $3 \times 10^6$\\
  $a$ (kpc) & 4.0 \\
 %\hline
 $b$ (kpc) & 0.1 \\
 %\hline
  $c$  & 12.0 \\
 %\hline
  $d$ (kpc) & 6.0 \\
 %\hline
  $r_{\rm vir}$ (kpc) & 258.0 \\
 %\hline
  $r_s$ (kpc) & 21.5 \\
  $c_{\rm sd}$ (\kmps) & 20.8\\
  $Z_{\rm disc}$ ($Z_{\odot}$) & 1.0\\
  $Z_{\rm halo}$ ($Z_{\odot}$) & 0.1\\
  $\rho_{\rm D0}$ ($m_p/cm^3$) & 3.0\\
  $\rho_{\rm H0}$ ($m_p/cm^3$) & $1.1 \times 10^{-3}$\\
 \hline
\end{tabular}
\label{table:setup-parameters}
\end{table}

\section{SIMULATION SET-UP}
\label{sec:setup}
Our initial set-up is quite similar to the one used by \citep[hereafter, S15]{KCSI}. The new feature that we explore is distributed star formation. The set-up comprises a high density disc embedded in a dark matter halo. The density distribution of the disc and halo gas, and consequently the pressure  distribution, has been derived using the combined gravitational potential of a stellar disc and dark matter halo.
	
\subsection{Gravitational Potential}
%The initial set-up of all the simulation results presented here is identical ({\ccr \bf{I mean identical to each other}}) (except where mentioned). 
  The potential due to stellar disc is described by the 
Miyamoto-Nagai potential \citep{Miyamoto1975}  in cylindrical coordinates, ($R, z$), as,
\begin{equation}
\Phi_{\rm disc}(R,z) = -\, \frac{G\,M_{\rm disc}}{\sqrt{R^2 + \left(a + \sqrt{z^2 +b^2}\right)^2}} \,\,\,.
\end{equation}
Here, $M_{\rm disc}$ is the total stellar mass in the disc. The parameters $a$ and $b$ represent the scale length and the scale height of the disc, respectively.
The dark matter halo has been modelled as a modified NFW profile \citep{Navarro1997} to reflect a finite dark matter density at the centre of the potential. The potential has been taken as,
\begin{equation}
\Phi_{\rm DM} = \frac{GM_{\rm vir}}{f(c)\, r_s} \frac{\log \left(1+ \sqrt{R^2 + z^2 + d^2}/r_s\right)}{\sqrt{R^2 + z^2 + d^2}/r_s}\,\,,
\end{equation}
where, $f(c) = \log(1+c) -c/(1+c)$, with $c = r_{\rm vir}/r_s$ as the concentration parameter, $r_{\rm vir}$ as the virial radius, $r_s$ as the scale radius of the NFW potential and $M_{\rm vir}$ is the mass enclosed within the virial radius. Here, $d$ represents the core radius of the dark matter halo. The exact values of the parametes are given in Table \ref{table:setup-parameters}.

\subsection{Setting up the gaseous distribution}
The initial gaseous distribution contains a high density, low temperature disc component which is enveloped by a low density, high temperature medium, called the circumgalactic medium (CGM). To keep the set-up realistic, the disc component has been made to rotate in the azimuthal direction. The hot component, however, is made to have zero rotational velocity for simplicity.  
Both of these components are in steady state equilibrium under the influence of the combined gravity, $\Phi = \Phi_{\rm disc} + \Phi_{\rm DM}$, of the stars and dark matter.

In such a situation, the steady state distribution for a component rotating with an azimuthal speed of $\rm{v}_{\varphi}$ can be described by
\begin{eqnarray}
\label{eq:HS-eq}
0 &=& - \frac{1}{\rho} \frac{\partial p}{\partial R} - \frac{\partial \Phi}{\partial R} + \frac{\rm{v}_{\varphi}^2}{R} \nonumber \\
0 &=& - \frac{1}{\rho} \frac{\partial p}{\partial z} - \frac{\partial \Phi}{\partial z}\,,
\end{eqnarray}
where, $\rho \equiv \rho(R,z)$ and $p \equiv p(R,z)$ are the density and thermal pressure, respectively. Now, for given $\rm{v}_{\varphi} \equiv \rm{v}_{\varphi}(R)$ and an equation of state, $p \equiv p(\rho)$, the above equations can be solved individually for both of the components. However, notice that we do not follow the evolution of the individual components but rather study them as a single fluid component. The density and pressure profiles for this single fluid are constructed from the linear superposition of the corresponding quantities of individual fluid components. This approach is effective at the gridpoints where any single fluid dominates the mass and energy budget of the cell (which is the case for most of the gridpoints) but fails to represent any physical quantity at the disc-CGM boundary where the contributions from both are comparable. This introduces certain artificial effects, \textit{viz} radiative cooling, at the disc-CGM boundary and requires some extra attention. A more detailed discussion to handle such artificial issues has been provided in S15.

As can be seen in equation \ref{eq:HS-eq} that for a test particle at $z = 0$, the gravity should be completely balanced by the centrifugal force which thereafter decides the rotational speed ( $\rm{v}_{\rm rot} = \sqrt{R\frac{\partial\phi}{\partial R}}\big\rvert_{z=0}$). However, for a realistic fluid, a fraction of the support against gravity also comes from the pressure gradient. We set this fraction by setting the rotation speed of the disc component to be $\rm{v_{\phi, g}} = f \rm{v}_{\rm rot}$. Here, we have chosen $f = 0.95$ for a realistic shape of the disc gas. This gives us for the disc gas,
\begin{eqnarray}
\rho_{\rm disc}(R,z) = \rho_{\rm D0} \exp \Big[ - \frac{1}{c_{\rm sd}^2} &\Big(&[\Phi(R,z) -\Phi(0,0)]  \nonumber \\
 - &f^2& [\Phi(R,0)-\Phi(0,0)]\, \Big) \Big] \,,
\end{eqnarray}
and for the halo gas,
\begin{equation}
\rho_{\rm halo}(R,z)=\rho_{\rm H0} \exp\left( - \frac{1}{c_{\rm sh}^2}\big[\Phi(R,z) -\Phi(0,0)\big]\right) \,.
\end{equation}
 Here, $\rho_{D0}$ and $\rho_{H0}$, respectively, are the central densities of the disc and halo gas, $c_{\rm sd}$ and $c_{\rm sh}$ are the isothermal sound speeds in the disc and the halo. This approach of finding the density and pressure distribution is quite similar to that used in \cite{Sutherland2007} and \cite{Strickland2000}. The sound speed in the disc has been taken be to be $24$ \kmps, which is slightly larger than the sound speed at $10^4$ K, to account for additional pressure from turbulence and non-thermal components. Observational evidence establishes that the central density of the disc, $\rho_{\rm{D0}}$, is close to $3$ m$_p$cm$^{-3}$. The central density of the hot gas has been estimated by normalising the total baryonic fraction in the galaxy to be equal to the cosmic baryonic fraction, $0.16$. Table \ref{table:setup-parameters} describes the parameters used in this paper.
% \tco{ It should be mentioned here that the above equations 3-5 have been derived in \citep{Sutherland2007} using a distribution function.}
 
 %
 \begin{table}
\caption{Identification of Different Runs}
\centering
\begin{tabular}{ || c|c| c | c | c|| } 
 \hline
 Name & Type & \texttt{$L_{\rm inj}$ (erg s$^{-1}$)} & $r_{\rm inj}$ (pc) & disc Resolution (pc)\\ 
 \hline
  \texttt{L42} & DI & $3.0 \times 10^{42}$ & 30  & 12\\
  \texttt{L41}  & DI &$3.0 \times 10^{41}$ & 20 & 12\\
 \texttt{L42\_CI} & CI & $3.0 \times 10^{42}$ & 60  & 20\\
 \texttt{L41\_CI} & CI &$3.0 \times 10^{41}$ & 60 & 16\\
 \hline
\end{tabular}
\label{table:2}
\end{table}

One issue, however, stays with treating the two components as one. In reality, the disc gas is inhomogeneous in nature and contains cold/warm clumps ($\sim 10^4$ K) along with a hot phase ($3\times 10^6$ K). Since we assume that the warm clumps are rotating while the hot phase is not, it brings us to a question - whether the hot gas will also be dragged by the warm clumps due to viscosity. We can estimate the time-scales over which such viscous force will be important.

For a $3\times 10^6$ K hot gas with a density of $10^{-3} m_p$ cm$^{-3}$, the kinematic viscosity is $\approx 5\times 10^{26}$ cm$^{-2}$s$^{-1}$ (see section $2.2$ of \citep{Guo2012}). Therefore, the viscous timescales to transfer the velocity of the warm clumps to a vertical layer of $l = 200$ pc is $t_{\nu} \sim l^2/\nu \sim 24$ Myr which is larger than the time-scale of our simulations. The Kelvin-Helmholtz time scale, for $100$ pc sized clumps, is of the same order. It is interesting to note that this process can transfer the disc rotation to the CGM to distance till $z \sim 4$ kpc over a time-scale of $10$ billion years. This may be a probable candidate to explain the vertical gradient of the rotation curve in spiral galaxies as observed by \cite{Heald2007, Kalberla2009}. However, a detailed study of such physics is out of the purview of the present paper.
  
\begin{figure*}
   \includegraphics[trim={0cm 0cm 0cm 5cm} ,clip=true, width=\textwidth]{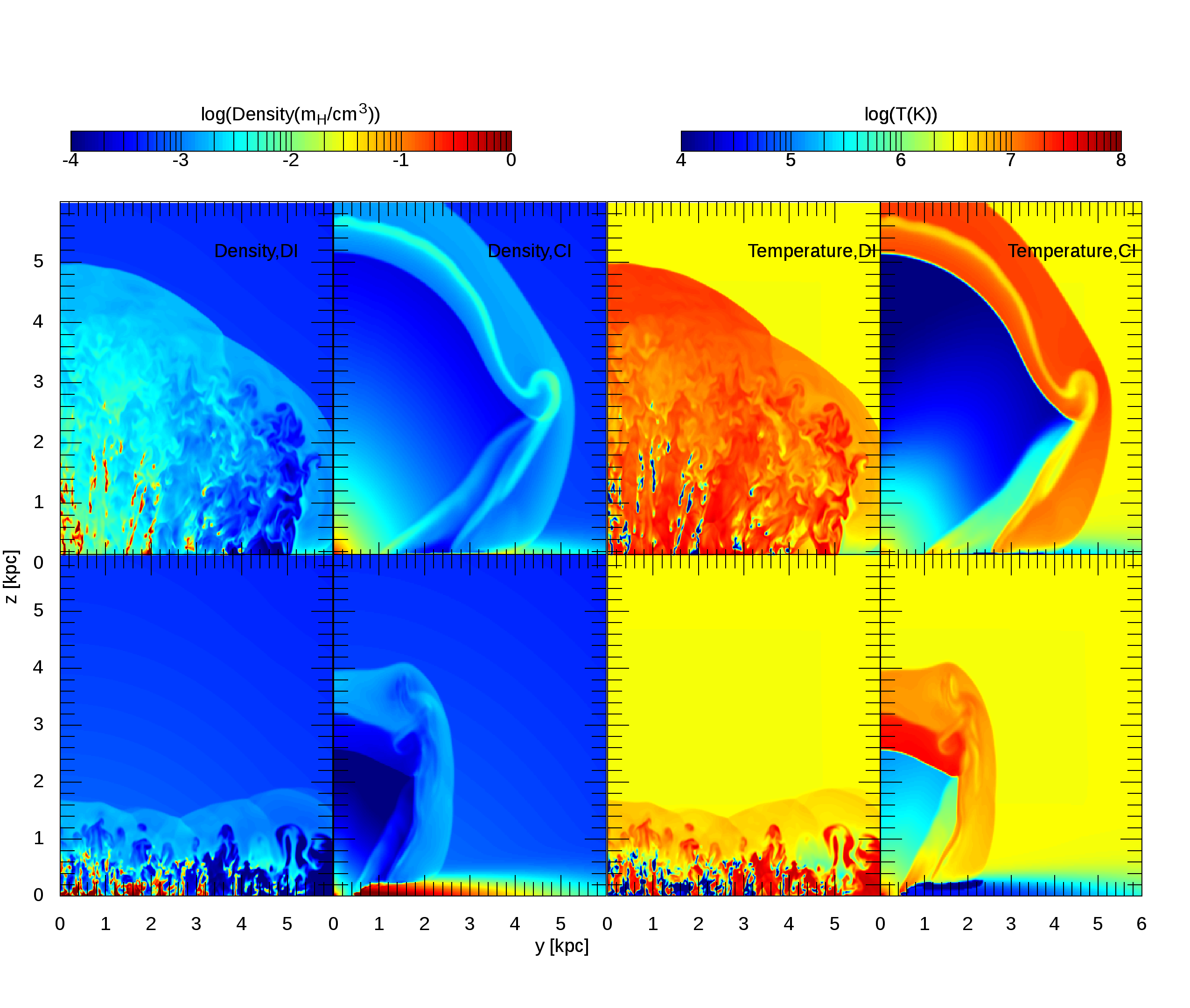}
 \caption{Density(left) and Temperature(right) contours for $x=0$ plane for $L42$ (upper panel) and $L41$ (lower panel) at $5$ Myr. The shock moves slightly faster in the case of CI, as compared to DI. Note that negative $y$-direction is shown here as it is morphologically almost similar to the $+y$ side, other than small differences.
 }
 \label{fig:densL42}
  \end{figure*}

\section{PARAMETERS OF THE RUN}
%{\ccr this section needs to be more elaborate-\bf{done...}}\\
Using the set-up described above we have conducted three-dimensional hydrodynamical simulations to mimic outflows resulting from star formation in the disc of the galaxy. The feedback due to star formation was replicated by a continuous injection of mass and energy in the computational domain. One purpose of this paper is to understand whether and how different aspects of the outflows, such as morphology and dynamics, change upon changing the modes of energy injection. In this respect, we would like to introduce the two different ways of injecting energy and mass into the system. The first of two, which will be referred to as Central Injection(CI) in the course of the paper, has been studied in S15. In S15, central injection was implemented in two dimensions with axisymmetry. The injection of mass and energy took place at the centre of the disc within a certain injection radius. Though this set-up faithfully reproduces the effects of outflows, a concentrated injection is not representative of the physically distributed star formation that takes place in a galaxy.

\begin{figure*} 
   \includegraphics[trim={0cm, 0cm, 0cm, 5cm}, clip=true, width=\textwidth]{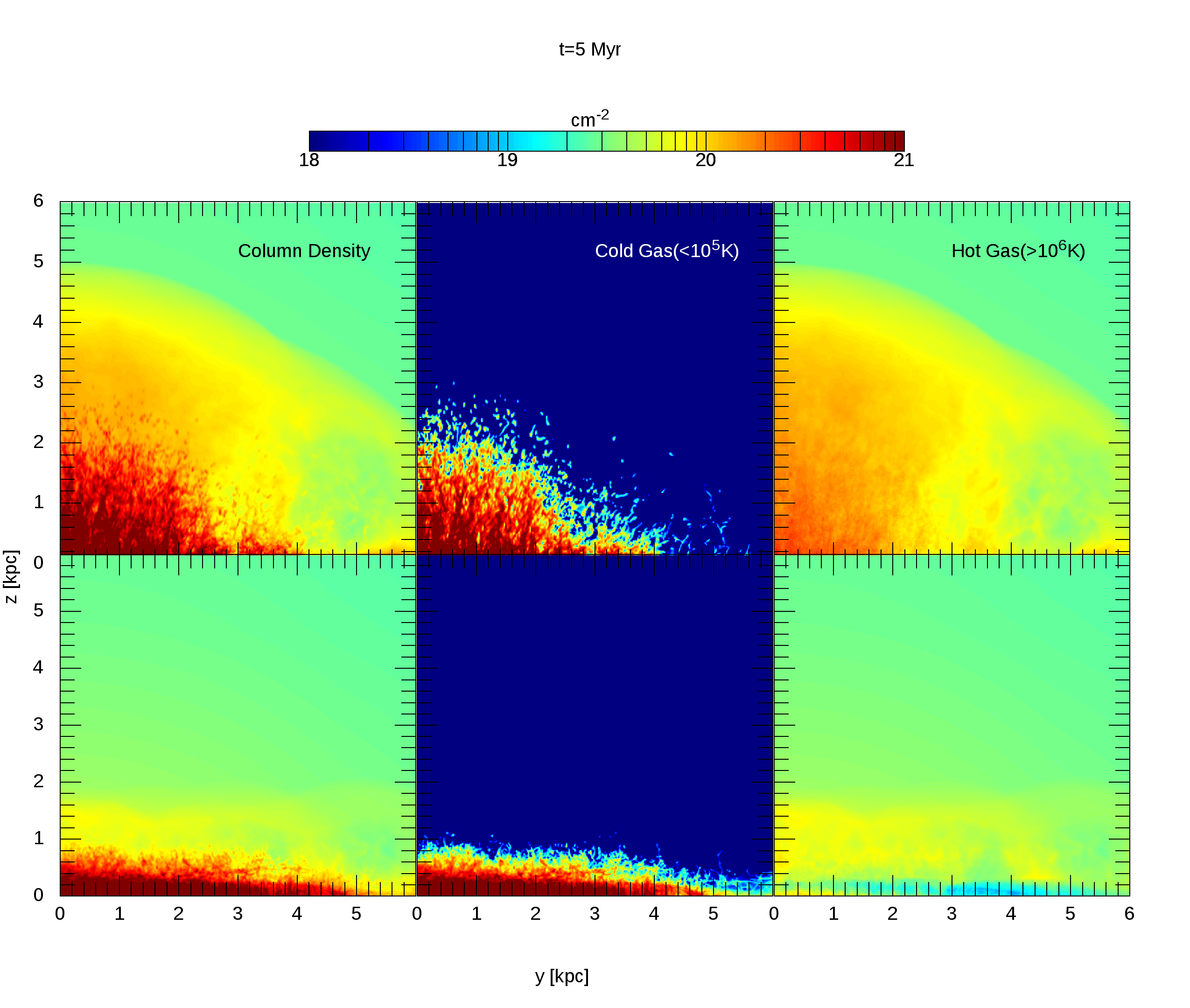}
 %  \includegraphics[width=\linewidth]{col_density_L42_10.png}
% \caption{Column Density Maps for L42 at 5 and 10 Myr.} 
%  \end{figure*}
  
%  \begin{figure*}
  % \includegraphics[width=\linewidth]{col_density_L41_05.png}
  % \includegraphics[width=\linewidth]{col_density_L41_10.png}  
 \caption{Column density maps for $L42$(upper panel) and $L41$(lower panel) at $5$ Myr. The number density of particles was added along the line of sight, that is $x$ direction. The filamentary structure of the warm (T<$10^5$K) as well as the hot gas is clearly visible in these maps. As can be seen, hot gas(T>$10^6$K) is diffused.} %and 10 Myr.} 
 \label{fig:columndensity}
  \end{figure*}

To capture the spread of star formation in the disc, we would like to introduce a mode of energy injection which we have christened Distributed Injection(DI). Instead of depositing the entire luminosity at the centre, we choose to have $1000$ points in the volume of the disc which represent sites of OB associations. 
Fewer injection points would result in an unrealistically large luminosity of each OB association, whereas larger number of points would require
an unreasonable computational power.
These $1000$ points will be referred to as injection points in the rest of the paper. The distribution of these points has been described in the appendix \ref{injection-points}.

All runs follow identical forms of potential and thus, initial density distribution, as in S15. Note that the CI set-up follow axisymmetry, unlike in the case of DI. For further details, the reader is directed to S15.

\begin{figure*}    
   \includegraphics[trim={0cm 0cm 0cm 5cm} ,clip=true, width=\linewidth]{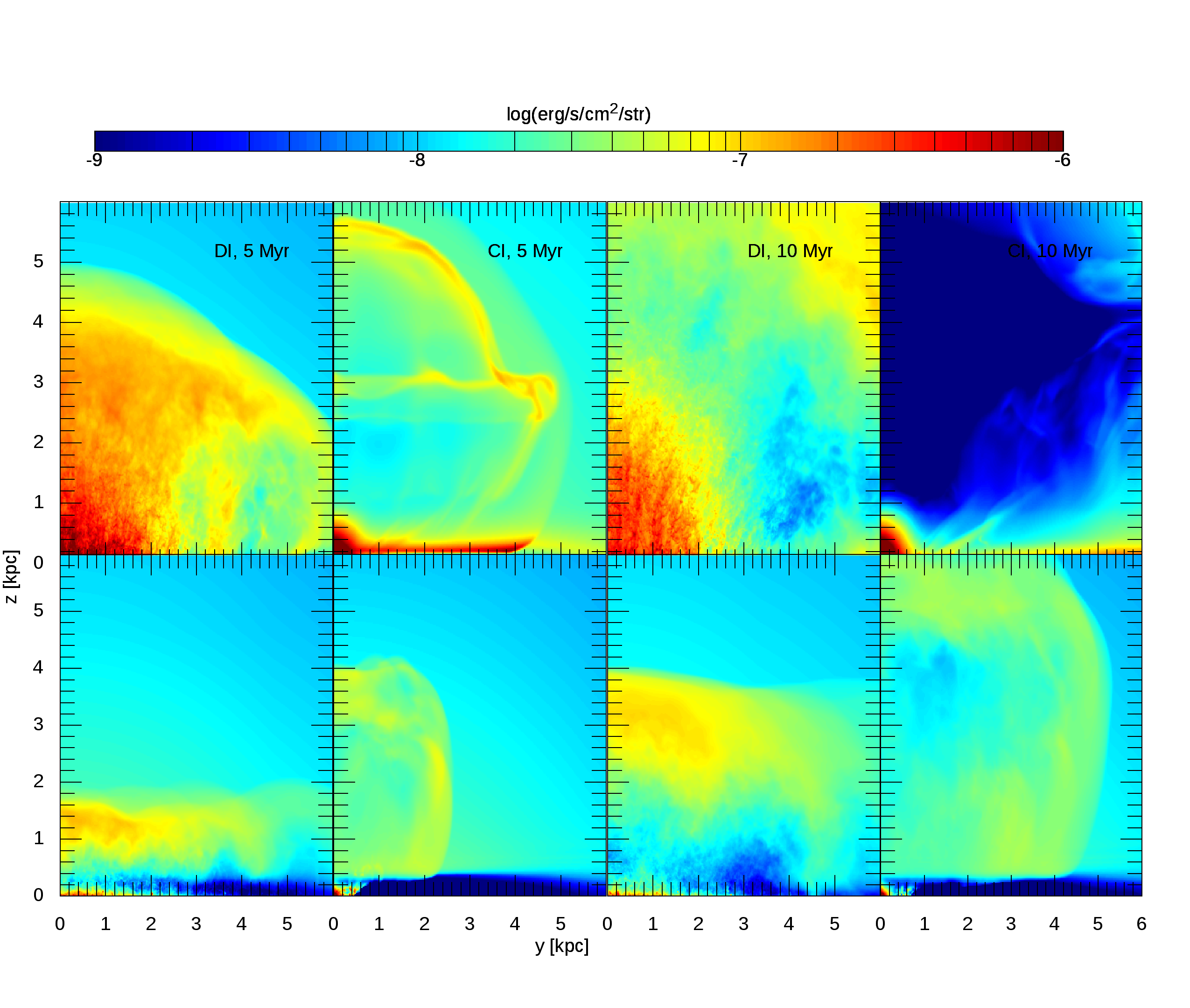}
 \caption{Soft X-Ray ($0.5\hbox{--}2$ keV) Surface Brightness($= \int_{\rm{LoS}} n^2 \Lambda dl /{4\pi}$) maps for $L42$(upper panel) and $L41$(lower panel) for DI and CI cases. We have defined $0.5$-$2.0$ keV as soft, here. Notice the filamentary structure of the emission visible in the case of DI. Also notice the cavity created as the shock moves out of the box in the case of CI. The halo contribution has not been taken into account for this map.}
 \label{fig:xraysb} 
  \end{figure*}

\subsection{Simulation set-up and injection points}
All the simulations mentioned in this paper were conducted using PLUTO.  PLUTO \citep{Mignone2007} is a finite volume code which solves the conservation laws,
\be
{\partial \bf{{U}} \over \partial {t}} = -\nabla \cdot T(\bf{U}) + \bf{S}(\bf{U}) \,,
\ee
where $T(\bf{U})$ is the flux tensor and $S(\bf{U})$ is the source term, which comprises $M_{\rm inj}$ and $E_{\rm inj}$ for mass and energy conservation, respectively.
We used the HD module offered by the code for all the simulations mentioned in this paper. 
The in-built cooling function in PLUTO has been used to account for cooling losses in the system. A temperature floor of $8000$ K was set. %({\ccr Describe PLUTO briefly here: finite volume code, can solve radiative losses using subcycling of timesteps.  describe where are the source terms are added in the fluid equatons. Or if possible, mention the equations.
%cite Mignone etal 2007})

The computational domain is a cuboidal box and uniform grid is applied in all directions. It extends from  $-6$ kpc to $+6$ kpc in the $x$ and $y$ directions and up to $6$ kpc in the $z$-direction, assuming symmetry about the $XY$ plane. Outflow boundary conditions are applied at all the boundaries except at $z=0$, where reflective boundary condition is imposed. The resolution of the runs is constrained by the energy injection parameters. As \cite{Sharma2014} suggested, the size of the region in which energy is deposited should be sufficiently small so as to prevent catastrophic loss of injected energy via artificial radiative losses. Geometry poses another constraint as spherical injection has to be implemented using cuboidal cells. This is overcome by ensuring that multiple (two to three in each direction) cells  occupy  the injection volume. The injected energy is completely thermalised.

The OB associations have been distributed throughout the volume of the disc such that surface density of star formation at a cylindrical distance $R$ in the plane of the disc is related to the gas surface density, integrated along the $z$-axis, by the Kennicutt-Schmidt law \citep{kennicutt1998}. The Kennicutt-Schmidt law thus dictates the number density of OB association along the $R$ axis of the disc. The distribution of OB associations consequently peaks at about $1.0$ kpc from the center (see equation \ref{nR}). Appendix A gives a detailed account of this implementation. It should be noted here that the off-center peak in the number of OB association is indeed has been corroborated by observations of the Milky Way \citep{higdon2013}. %Interesting is the fact that this arises purely as result of the cylindrical geometry of the disc and is a direct consequence of the Kennicutt-Schimidt law.

The distribution of these  $1000$ OB associations along the $z$-axis has been determined by the local density scaled with the local mid-plane density. Further, the OB associations are rotating with the local circular speed in the disc and trace circular orbits.
The injection points are surrounded by a suitably chosen injection region in which energy and mass are deposited continuously throughout the run period, that is $10$ Myr, in a low density medium. The runs have been conducted for two values of the total energy deposited (referred to as $L42$ and $L41$ subsequently) in the disc, which have been distributed equally over the $1000$ injection regions. The mass injection rate has been estimated using the Kroupa mass function. Consequently, the energy and mass injection rates are related to the star formation rate(SFR, in $M_{\odot}$ yr$^{-1}$) as,
\be 
L = 10^{41} \, {{\rm erg} \over {\rm s}} \, \Bigl ( {{\rm SFR} \over {\rm M}_\odot \, /{\rm yr}} \Bigr ) , \qquad
\dot{M}_{\rm inj} =  0.1 \, {\rm SFR} 
\ee

\subsection{Tracers}
PLUTO allows for the evolution of advection equations, which have been used to trace the disc and injected matter, as well as the halo gas, over the period of the runs. Initial density profile was used to set the tracers for the  disc gas and the halo gas. The injection region was treated as source of the tracer for the injected material. Different regions have different tracers that are advected by the flow.

\begin{figure*}
	\centering
	\includegraphics[width=0.9\textwidth]{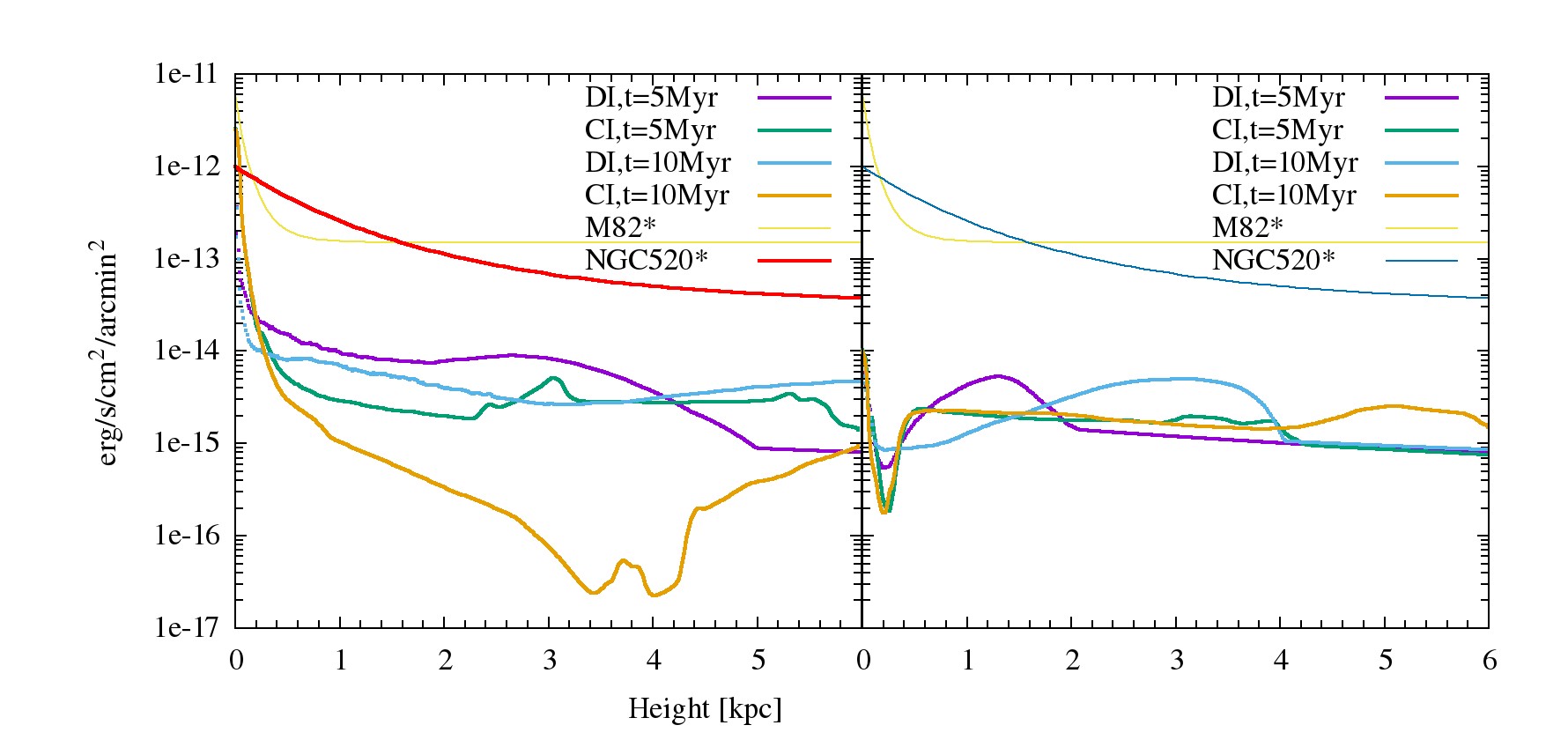}
	\caption{Average X-Ray VBD (see equation\ref{VBD}) along the z-axis, starting from $z=0$ for DI, CI for at $5$ and $10$ Myr, for $L42$ (left) and $L41$ (right) cases. The emissivity drops more sharply for CI case as compared to DI case. The bumps in the left figure represent the location of the shocked shell. The arrested fall of the emissivity compares will the observations(read text for details). For comparison with observation, it can noted that for typical numbers $1\times 10^{-13}$erg s$^{-1}$ cm$^{-2}$(emitted by plasma of solar metallicity) corresponds to $6.6\times 10^{-3}$ counts/s for Chandra ACIS-S $0.5-2.0$ keV. This estimate does not include the contribution by the CGM which may be of $\sim 1.5\times 10^{-14} n_{10^{-4 \rm{cm}^{3}}} \Lambda_{10^{-23}}$erg s$^{-1}$ cm$^{-2}$ arcmin$^{-2} $. *The plots have been made using scaleheights for exponential fits of figure 5 in \citep{Li2013}. } 
	\label{fig:xray1D}
\end{figure*}

\section{RESULTS}
\subsection{Density and temperature}
Density and temperature contours for DI and CI cases for $L=3 \times 10^{42} \rm{erg s^{-1}}$ (corresponding to SFR=$30$ M$_\odot$ yr$^{-1}$) are shown in Figure \ref{fig:densL42}, for the $x=0$ plane at $5$ Myr. The contours show crucial differences that translate into differences in the X-Ray surface brightness maps, as discussed below. As a result of the initial density profile, the CI case exhibits strong symmetry about the $R$ axis of the disc and the outflow displays the characteristic shock structure, with distinguishable reverse shock, contact discontinuity and forward shock, as expected in the classic stellar wind model \citep{Weaver1977}. However, for the DI case, such a straightforward demarcation is nearly impossible, particularly in the inner region of the disc, where the SFR is high. The free wind region in DI is occupied by warm clumps of disc gas (discussed later), whereas in CI, this
region is devoid of clumps. Previous works (e.g., S15) have noticed that the mass beyond the contact discontinuity fragments due to thermal  and later Rayleigh-Taylor instability. In our case, for CI, the contact discontinuity has not formed clumps owing to the rather long cooling time ($\sim 60$ Myr, given temperature of $\sim 5\times 10^6$ K and density of $\sim 5 \times 10^{-3} $ m$_p/$ cm$^{-3}$ in the shocked ISM region). Moreover, at a very early stage, when the shell is highly radiative, the diverging velocity field in the shell inhibits any clump formation. %These clumps in the free wind region develop a bow shock which  contribute significantly to the X-ray emission (see below). 
It should be added this point that for $L42$, both DI and CI, the shock moves out of the computational domain after a period of $5-6$ Myr.
The speed of the shock varies marginally in the two cases. In the CI case, in the approximation of self-similar evolution of the forward shock, the radius and speed of the outer shock through a uniform medium are
given by (for energy input rate $L$ and ambient density $\rho$),
\be
R_{CI}\approx \Bigl ( {L t^3 \over \rho} \Bigr ) ^{1/5} \,, \qquad V_{CI} \approx {3 \over 5} \Bigl ( {L  \over \rho} \Bigr ) ^{1/5} t^{-{2/5}} \,.
\ee
In the distributed case, one can roughly assume a parallel slab geometry for the central region. The relevant parameters here are time ($t$), ambient density ($\rho$)
and the surface density of energy input, $\Sigma_L=L/(\pi R_d^2)$, where $R_d$ is the radius of the region of star formation. These parameters can be combined to give the height and speed of the forward shock, as,
\be
R_{DI} \approx \Bigl ( {\Sigma_L t^3 \over \rho} \Bigr ) ^{1/3} \,, \qquad V_{DI} \approx \Bigl ( {\Sigma_L  \over \rho} \Bigr ) ^{1/3}  \,.
\ee
This is valid until the outflow geometry tends to become
spherical, when $R_{DI} > R_d$.
The ratio of the shock speeds in the two cases is,
\be
{V_{CI} \over V_{DI}} \approx {3\over 5} \Bigl ( {L \over \rho} \Bigr ) ^{-2/15} \, t^{-2/5} \, \Bigl (\pi R_d^2 \Bigl ) ^{1/3} %\nonumber\\
\approx  L_{42} ^{-2/15} \, \rho_{-3}^{2/15} \, t_5^{-2/5} \, R_{d,4}^{2/3} \,,
\label{eq:ratio-speed}
\ee
where $L=L_{42} 10^{42}$ erg s$^{-1}$, $\rho= \rho_{-3} 10^{-3} \, m_p$ g cm$^{-3}$, $t=t_5 \, 5$ Myr, and $R_d= R_{d,4} 4$ kpc.

From our simulations we find that the forward shock for $L=3 \times 10^{42} \rm{erg s^{-1}}$ reaches $\sim 5$ kpc and slightly more than $6$ kpc in $5$ Myr in the CI 
and DI cases, respectively. The comparable heights reached in the two cases are consistent with the above estimate, given the approximate nature of the estimate. 
For a lower SFR, with $L=3\times 10^{41}\rm{erg s^{-1}}$, the corresponding heights are $\sim 4$ and $\sim1.8$ kpc, showing that the ratio of heights in these two
cases increases as luminosity decreases, again consistent with the scaling derived above, within factors of order unity.

We note that the shock speed in the DI case remains roughly constant with time, whereas in the CI case the shock decelerates. The ratio of the speeds  ($V_{DI}/V_{CI}$) in the case of $L42$ is roughly unity at $\sim 5 $ Myr, but it increases with time. In other words, for a longer
time scale, the shock speed in the distributed case is expected to be larger than in the central injection case. This is an important difference, owing to the geometry
in two cases, and we will discuss the implications below.
Ultimately, after the outer shock becomes larger than the injection radius, the plane-parallel case should transform into a spherical wind.
 
%A crude calculation tells us that the forward shock, the only discernible feature of shock structure in $x=0$ plane, moves with a speed of $1000$ \kmps, as it has reached a height of $5$ kpc in $5$ Myr for DI, figure \ref{fig:densL42}. The forward shock moves faster in CI, covering slightly more than $6$ kpc in the same time. The average speed can be taken to be $1200$ \kmps. %The speed of the shock can related to the underlying parameters of the system such as the surface density of the disc and the extent of the energy injection(see appendix for scaling relations).

We show the corresponding column density maps in Figure \ref{fig:columndensity}. We integrated the density along the $x$-axis to determine the column density.
Further, we have distinguished between the column density of hot and warm gas, as shown in the middle and right panels of the figure. Gas with a temperature less $10^5K$ is taken as warm gas, while that which is hotter than $10^6K$ is taken as hot gas. The figure shows that warm gas is thrown up to height of $\sim 2$ kpc in $5$ Myr, while hot gas has reached nearly the edge of the box at $\sim 6$ kpc.

%\subsection{Column Density Maps}
%\begin{itemize}
%\item Density is integrated along the x-axis and the column density is arrived at.
%\item Gas with a temperature less $10^5K$ is taken as cold gas, while those parcels with $10^6K$ are taken as hot gas parcels.
%\item As is evident from the figure 1, cold gas is thrown up to height of 2kpc in 5Myr, while hot gas has reached nearly the edge of the box at 6kpc.
%\item It should be pointed out that X-Ray emission maps do not have the clumpy nature that are exhibited by the cold gas maps but seemed to be diffused like the hot gas maps.
%\end{itemize}

\begin{figure*}
	\centering 
	\includegraphics[trim={0 0 0 2cm}, clip=true, width=0.9\linewidth]{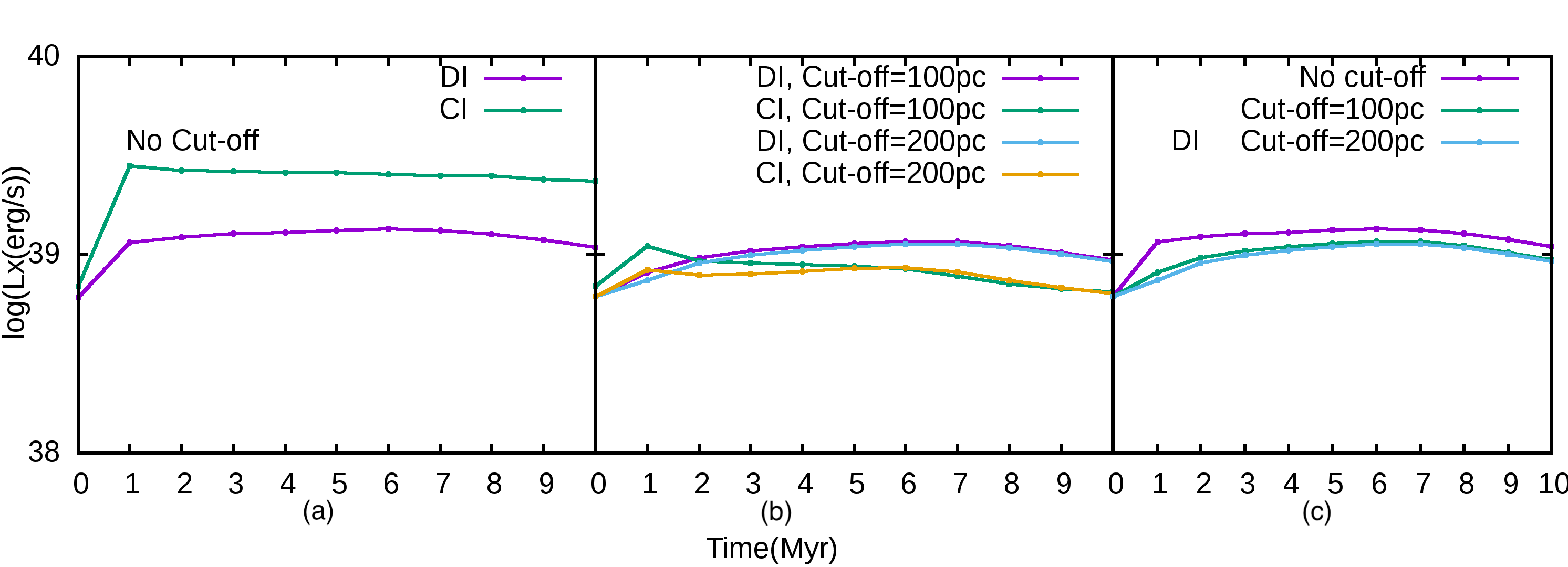}
	\caption{From left to right-(a) Total $L_{\rm{xs}}$ without cut-off;(b) Cut off of 100 pc and 200 pc;(c) A comparison between total $L_{\rm{xs}}$ for different cut-offs. Although, the total soft X-Ray luminosity is higher for CI, once the contribution from the injection region is removed DI has higher lumniosity. For CI, highest contributor to the X-ray luminosity, after the injection region, is the shocked shell. For DI, however, the bow shocks occupying the free wind region also contribute significantly. The cut-off has been taken along the $z$ axis, across the horizontal plane. $L41$ would be dominated by X-ray emission from the halo owing to a low SFR, as concluded by \citep{Sarkar2016}(their Figure 3).} 
	\label{fig:totlx100200}  
\end{figure*}

\subsection{Soft X-ray emission}
\label{sec:soft-xray}
We show the soft X-ray ($0.5-2$ keV) surface brightness maps in Figure \ref{fig:xraysb}, for $L=3 \times 10^{41}$ erg s$^{-1}$ and $L=3 \times 10^{42}$ erg s$^{-1}$, at $5$ and $10$ Myr, for both CI (central injection) and DI (distributed injection) cases. Qualitatively, CI and DI produce morphologically different surface brightness maps. The most important difference between the two cases is that in the CI case the X-ray surface brightness map shows a central region and a shocked shell as discussed in the previous section. In this case, the X-Ray luminosity is dominated by gas in a symmetric shell that expands with time. After nearly $10$ Myr, the forward shock and contact discontinuity have moved out of the computational domain, leaving behind a cavity-like structure in the surface brightness map owing to the cooled free-wind region.

In contrast, the emission in DI case is dominated by filament-like structures that fill the space between the disc and the outer shock (whose brightness is less than in the case of CI). The filamentary structure is the result of the interaction of the warm clumps of disc gas with the free wind. The X-Ray luminosity consequently decreases smoothly as a function of height from the disc, as opposed to a sharp decrease in the case of CI. This is consistent with the findings of \citep{Li2013} according to whose observations the  vertical brightness distribution(VBD),
\be
\label{VBD}
\rm{VBD} = \frac{\int n^2 \Lambda \rm{dx}\rm{dy}}{\int \rm{dy}}
\ee

 ,along the minor axis of star forming spiral galaxies is such that it decreases by a factor of few tens from the central peak value to the value for the coronal gas far outside. We have plotted the VBD of the soft X-ray along the line of sight, $x$ axis, against the height from the disc in Figure \ref{fig:xray1D} for both CI and DI. The drop in the distribution is much steeper for CI, which is expected since the injection region is the strongest contributor to X-ray followed by the shell. Note that the discrepancy between the observed profile and the obtained profile can arise due to several reasons. Firstly, while we average the VBD over the entire box, the observations use only the X-ray bright regions to find the averaged brightness. Secondly, our VBD  considers only a small box of size $-6$ to $+6$ kpc in X, Y and $0$ to $6$ kpc in Z direction, whereas, there may be some contribution from the hot CGM in reality. Lastly, incorporation of thermal conduction may affect the actual value of the brightness as we expect higher density of the hot gas due to evaporation of the clumps. % more clumps and therefore more bow shocks to form at higher resolution which can increase the surface brightness (see appendix \ref{appendix:resolution-study}).

A comparison of the total soft X-ray luminosity ($L_{\rm{xs}}$) in both the CI and DI cases has been shown in Figure \ref{fig:totlx100200}. It shows that $L_{\rm{xs}}$ sharply rises above the contribution from the CGM ($L_{\rm{xs}}$ at $t=0$) and then remains  almost constant for $t \ge1$ Myr. This sharp rise occurs due to the injection of the thermal energy in the disc. The rise time is roughly the time-scale over which a steady wind structure is set up within the injection regions. This time-scale is $\approx R/\sqrt{\dot{E}/\dot{M}} \approx 1$ Myr. Here, $R$ is the size of the injection region, $\dot{E}$ and $\dot{M}$ are the constant energy and mass injection rates.

We notice that if the entire computational domain is included, CI has marginally greater soft X-ray luminosity than DI (as shown in the left panel of figure \ref{fig:totlx100200}). However, as previous works have shown
(e.g., simulations of S15, or theoretical estimates of \cite{Zhang2014}), the X-ray luminosity the outflowing gas is dominated by the central injection region. This contribution to the soft X-ray luminosity is liable to be obscured by the dense gas in the central region. Therefore we need to compare the luminosities in the two cases besides this central region. For this, we exclude the disc up to $100$ and $200$ pc height. Once this region is excluded, we find that the reverse holds, and 
the luminosity for the DI exceeds the CI case (second and third panel of Figure \ref{fig:totlx100200}). This demonstrates again that for CI, the injection region is the strongest contributor to the X-ray luminosity, followed by the shock. Since the total energy deposited in both the cases is identical and the shock speed in the two cases is comparable (see previous section), it can be safely assumed that luminosity of the shock of one is within an order of the other. The difference thus arises from other sources. As we explain below, this excess in DI is supplied by the bow shocks of the warm clumps coexisting with the free wind. These bow shocks are absent in the CI case due to lack of the warm clumps in this scenario, because the region of interaction between the injected and the disc gas is limited compared to the DI case.
%Further,  
%Total X-ray luminosity for DI is marginally above that for CI. However, if the volume is limited to discount the contribution of the disc the trend reverses, as in figure \ref{fig:totlx100200}. The change in $L_x$ is obviously dependent on the cut-off chosen. %A comparison is shown in Figure \ref{fig:totlx100200}.

\begin{figure*}
\centering
\includegraphics[width=\textwidth]{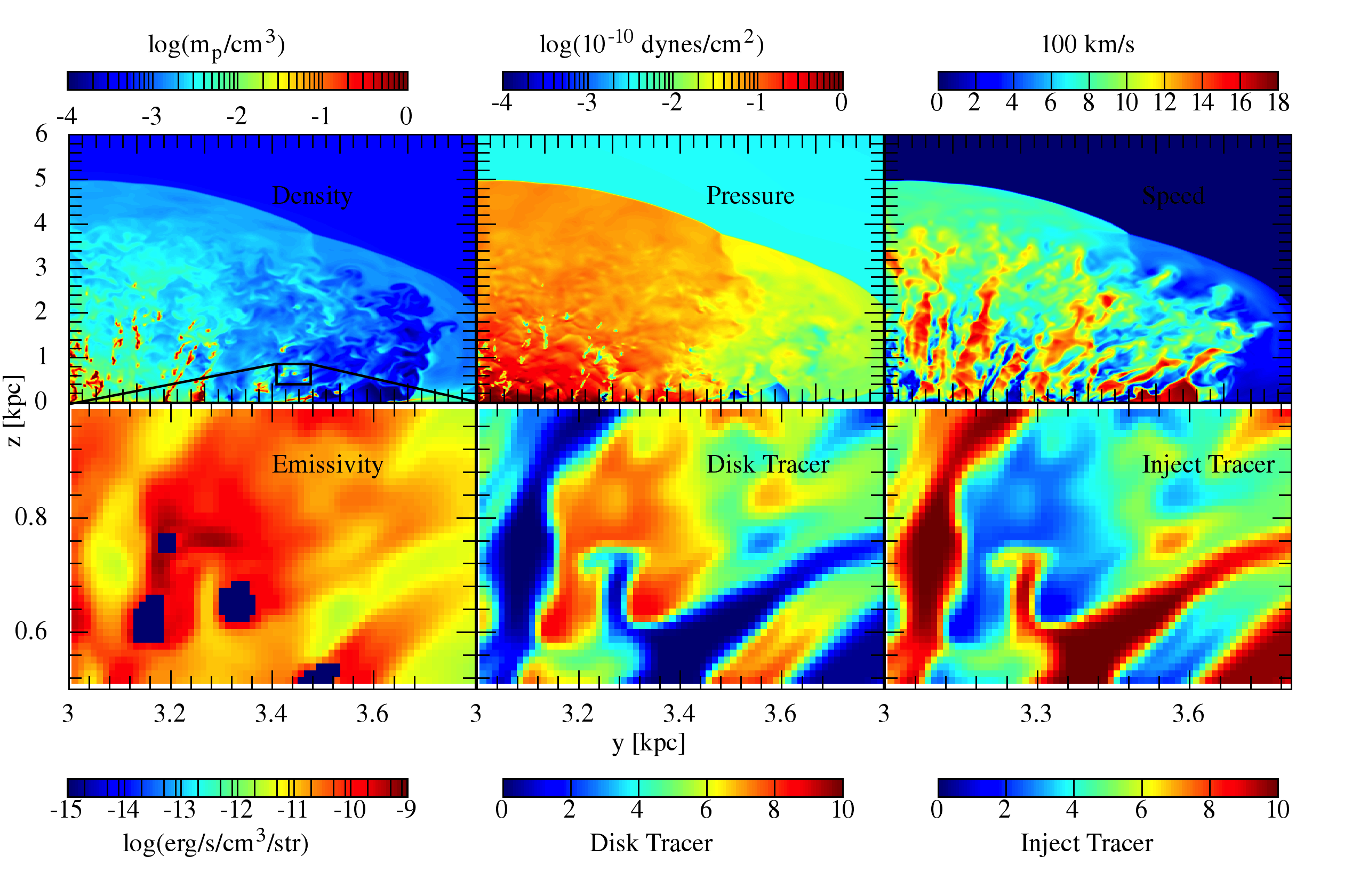}
	\caption{Density, pressure and speed($\sqrt{v_x^2 + v_y^2 + v_z^2}$) (top panel) and the emissivity, disc and injected tracer (bottom panel) contours in the $x=0 $ plane for $L42$ at 5 Myr. (The section of the density plot, shown inside the black box, has been zoomed into for emissivity and tracer plots.) The disc tracer and the injected gas tracer follow the advection equation for the disc gas and the injected gas, respectively. Numerically, they are dimensionless quantities, set between 0.0 and 10.0 having been pegged to the density of the material being traced. The parcels with a higher value of the disc(injected gas) tracer have a higher amount of disc gas(injected material). The clumps as seen in the density contour are predominantly comprise disc gas, embedded low density injected material. These warm clumps have negligible soft X-ray luminosity due to low($\sim 10^4$K) temperature, but the intervening region has high emissivity, since its temperature lies between that of the clumps and $\sim 10^{6-7}$K of the surrounding medium. (Figure \ref{fig:densL42} shows the temperature map.)} 
	\label{fig:zoom}
\end{figure*}
  
\subsubsection{X-ray from bow shocks}
\label{subsubsec:bow-shock}
One of the important questions we would like to address about the X-Ray luminosity of star forming galaxies is the source of X-Ray emission. It has been noticed in previous works that the impact of a warm clump against hot, high speed winds leaves a wake of hot gas behind it \citep{Cooper2008, Cooper2009}  which can be the dominant source of X-Rays given the temperature and the enhanced density. In this regard, it is worthwhile to notice that CI produces nearly no clumps because of the large cooling time (as discussed in section 4.1) of the shocked ISM and a strongly divergent velocity field. We zoom into one of the clumps in Figure \ref{fig:zoom}. The two left panels show the density and soft X-ray emissivity contours (top and bottom, respectively), and the bottom two panels on the right show the contours of the disc and injected gas tracers. It is clearly seen that bow shocks are formed as a result of the relative motion of the clump and the free wind. These bow shocks 
have high X-ray emissivity arising from high relative velocity ($\sim 500$ \kmps) and high density arising from the interface between the dense warm clumps and free wind. Note that the density of such bow shocks may be dependant on the spatial resolution of the simulation and will be discussed later.

\begin{figure}
   \includegraphics[width=7cm, height=7cm]{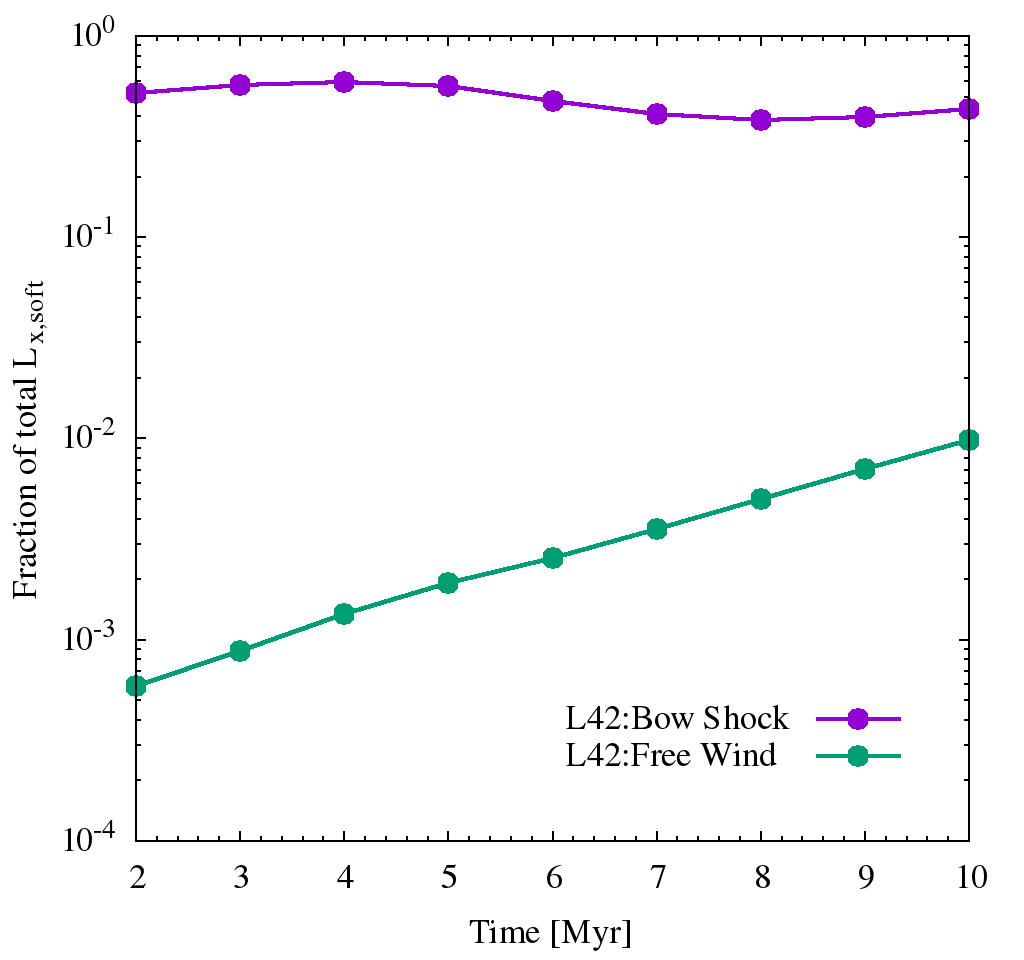}
 \caption{Fraction of X-Ray luminosity in bow shocks and free wind against time for $L42$, for DI. For this plot, it was assumed that free wind contribution comes from gas parcels with speed more than $1500$ \kmps. (The contribution of the injection region was removed by applying a cut-off of 100pc in the $z$ direction.) The bow shock contribution has been estimated by assuming that after a certain point the shocked shell exits a smaller box of size $8 \times 8 \times 4 $ kpc$^3$. Since contribution from the injection region and the shocked shell has been removed, we can conclude that a significant amount of X-Ray luminosity is sourced from the bow shocks. %Note that, the contact discontinuity in case of L41 has not yet cleared the box at this time.
 }
 \label{fig:totLxfwbs}
  \end{figure}

One can ask the relative contribution of these bow shocks to the  total X-ray luminosity of the outflowing gas. For this purpose, we selected a box 
 $-4$ kpc $< x,y < +4$ kpc and $100$pc $< z < 4$ kpc, 
%$8\times 8\times 4$ kpc$^3$ (with $4$ kpc height) 
around the central region, which contains most of the clumps and bow shocks at $5$ Myr. In this box, there is also the free wind region apart from the clumps and bow shocks. Since we know the speed of the free wind material, we identify this region and determine its X-ray luminosity. The rest of the 
X-ray luminosity arises from the bow shocks, because the contact discontinuity and consequently the shocked CGM shell has moved out of the $8\times8\times 4$ kpc$^3$ box by $5$ Myr. We show in Figure \ref{fig:totLxfwbs} the fraction of the total X-ray luminosity (in the whole simulation box) that is contributed by the free wind and the bow shocks, after a time scale of $5$ Myr, for the
$L41$ and $L42$ cases.

The figure shows that as much as half of the total X-ray luminosity of the outflow arises from the bow shocks, and this fraction remains roughly constant 
with time. This is one of the most important results of our paper. Note that the curve for the bow shock fraction for $L42$ case decreases slightly after $5$ Myr, due to 
the fact that after that time bow shocks begin to move out of the selected $8\times 8\times 4$ kpc$^3$ box.

The origin of these clumps is another aspect we would like to address. The clumps could comprise mostly disc gas which is blown away by continuous energy injection. The two bottom right panels of Figure \ref{fig:zoom} show that the clumps comprise of disc material and not the injected matter, since the clump regions contain
matter with the highest values of disc tracer. The disc and the injected tracer in the two right most panels of the figure, unambiguously, show that clumps are essentially disc gas embedded in a medium of the injected gas. It should be noted here that \cite{Cooper2008} attributed the clumps to initial density perturbations in the disc, arising out of turbulence. However, our result demonstrates that clumps form even if the disc gas is homogeneous but stratified, according to the
hydrostatic equilibrium, in the gravitational potential because of thermal and Rayleigh-Taylor instabilities.

\begin{figure}
   \includegraphics[trim={0 0 0 1.5cm}, clip=true, width=8cm,height=6cm]{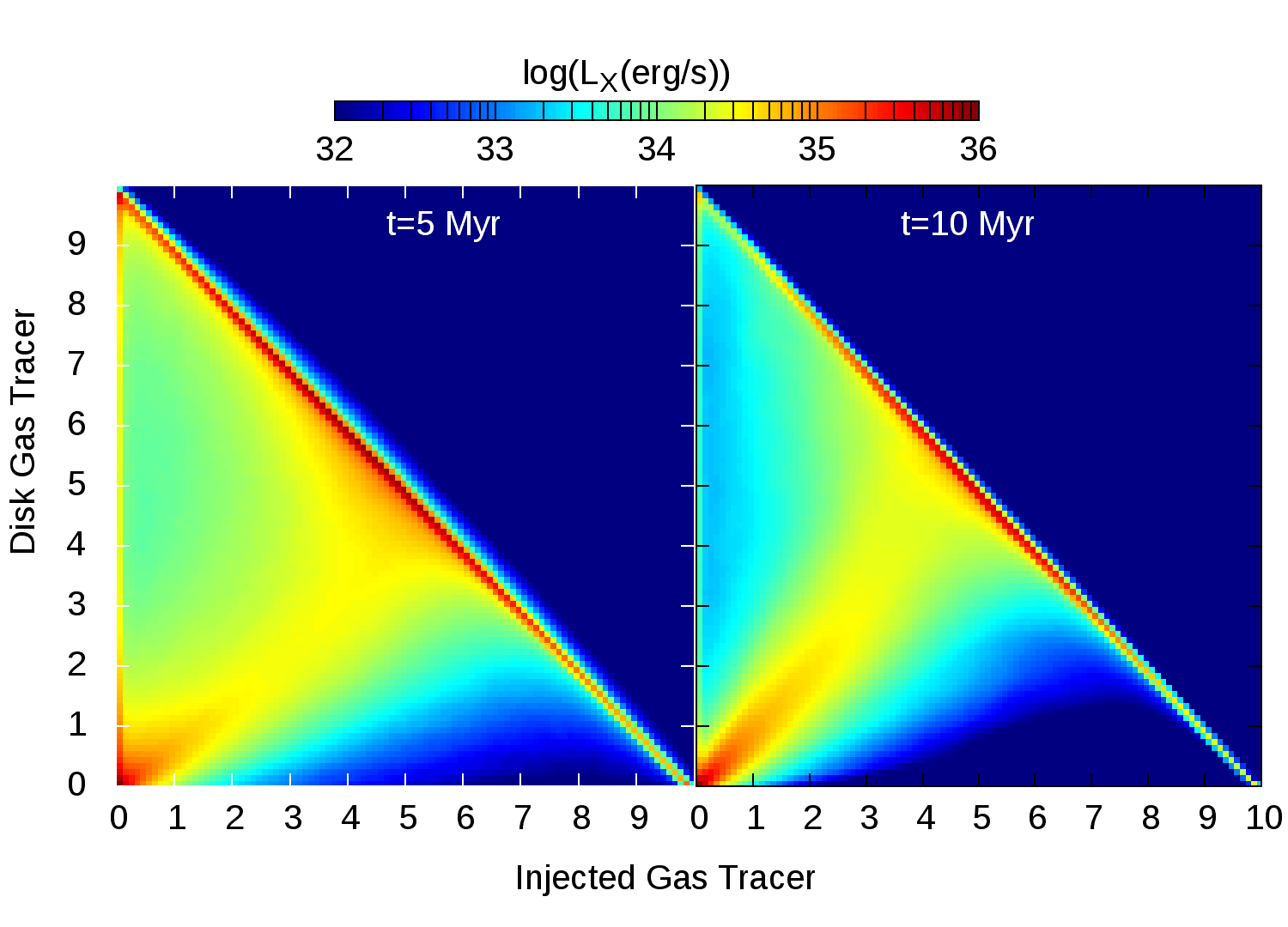}
 \caption{Luminosity-Tracer Maps for $L42$ at $5$ and $10$ Myr. These maps depict how much of the soft X-ray luminosity is contributed to by the disc gas and the injected material. The highest luminosity region in this map, the diagonal from top-left to bottom-right, represents those regions where there has been mixing between the disc gas and the injected material. As explained in the text, these are the regions around the clumps(see Figure \ref{fig:zoom} and explanation therewith) with intermediate temperatures. The bottom left corner depicts the contribution from the halo. (This figure includes contribution from $z>100pc$.)
 }
 \label{fig:lumtrc}
  \end{figure}

 We also show, in the two panels of Figure \ref{fig:lumtrc}, the parameter space of disc and injected gas, colour coded by the X-ray luminosity, at 5 and 10 Myr for $L42$ . %A parcel is taken to be mostly of disc gas if the disc gas tracer is more than $7.0$ and the injected gas tracer is less than $3.0$.
The bright bottom left corner tells us that the hot halo gas, which will by design have low disc and injected material tracer, contributes to the X-ray luminosity. We also note that gas which  solely arises from the injected material (along the horizontal axis) does not contribute much to X-Ray luminosity.
  The other distinctive feature of these maps is the bright diagonal from one end of the plot to the other, which is conspicuously brighter near the middle. The diagonal represents those parcels of gas in which the clumps of disc material have mixed with the injected material and are devoid of halo gas. The regions in which the halo gas mixes with disc and injected gas are the sub-diagonal points in this figure. This observation is further strengthened upon comparing the maps at two different times. At $10$ Myr, the shocked shell has left the computational domain and hence does not contribute to the X-ray luminosity. All that is left is the free wind region in which the disc gas in the form of clumps is mixed with the injected material.
  The bright diagonal further corroborate the observation that the bow shock regions, which are formed at the interface of disc-material-enriched clumps and
 injected-material-enriched free wind gas, are the dominant contributors to the X-ray luminosity. 
  \cite{Cooper2008} hypothesized that the bow shocks  formed as a result of the impact of the wind against the clumps are an important source of soft X-rays, although a quantitative analysis was not provided (see their Figure 15).

  \subsection{Hard X-ray emission}
  As the estimate in equation \ref{eq:ratio-speed} suggests, the geometry of the distributed star forming regions in a disc causes the speed of the shock
  to remain roughly constant, while that in the CI case decreases with time. If we discount the central region, which emits harder X-rays than the soft X-ray
  range considered here ($0.5\hbox{--}2$ keV), 
  the hardness ratio ($L_{x, \, >2\, {\rm keV}}/L_{x, \, 0.5\hbox{--}2 \, {\rm keV}}$ ) is expected to remain constant in DI case whereas the X-ray emission is expected to become softer for CI case.
  
%  X-ray emission is expected to become harder with time in the DI case compared to 
%  the CI case. 
  
 \begin{figure*}
 \includegraphics[trim={0 0 0 2cm}, clip=true, width=\linewidth]{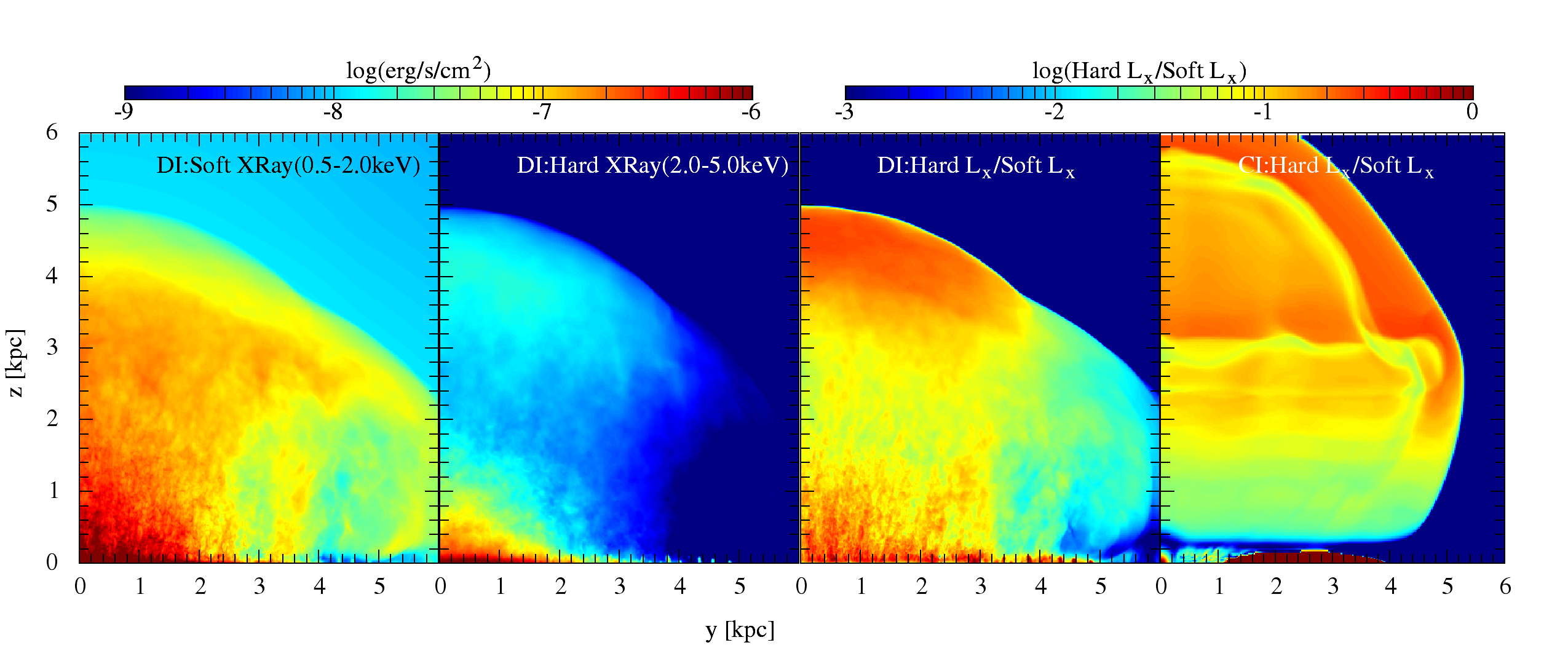}
  \caption{From left to right-(a) Soft X-Ray($0.5-2.0$ keV) surface brightness for DI for $L42$ at $5$ Myr; (b)(a) Hard X-Ray($2.0-5.0$ keV) surface brightness for DI for $L42$ at $5$ Myr; (c) Hardness ratio, i.e., ratio of the hard X-ray to the soft for DI for $L42$ at $5$ Myr; (d) Same as (c) but for CI. 
 }
 \label{fig:hardness-contour}
 \end{figure*}

\begin{figure}
\centering
\includegraphics[width=\linewidth]{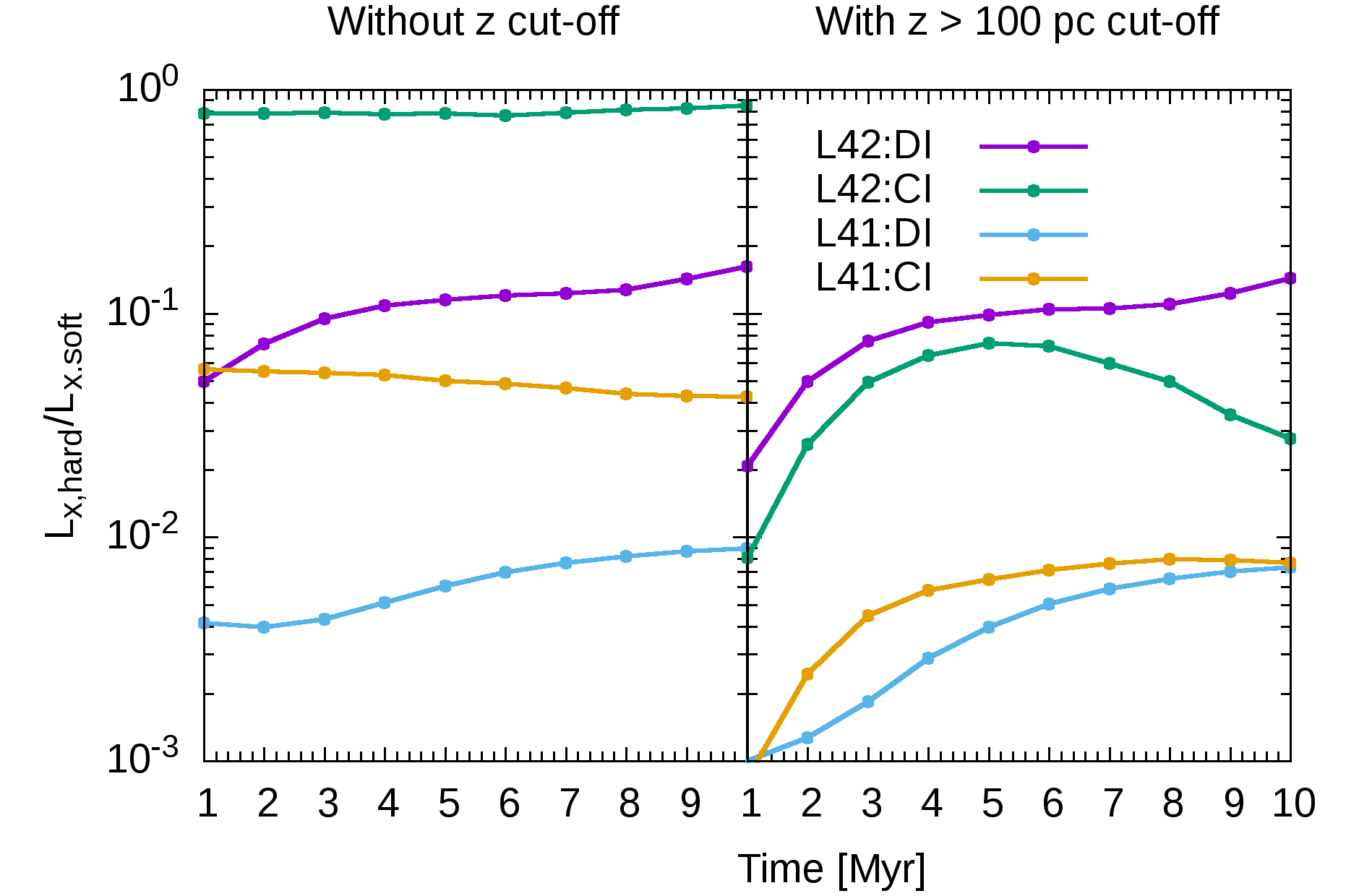}
 \caption{The ratio of hard X-Ray($2.0-5.0$ keV) luminosity to the soft X-Ray luminosity has been plotted against time. In the right panel, we have removed the contribution from the injection region by imposing a cut-off of 100pc along the $z$ direction. As is expected, this ratio decreases significantly when the cut-off is imposed since the source region is a strong contributor of hard X-Rays. But the overall contribution from the rest of the volume is non-negligible.}
\label{fig:totLx_soft_hard}
\end{figure}
  
  We compare the soft and hard X-ray surface brightness for the DI case ($L42$, $5$ Myr) in the two left panels of Figure \ref{fig:hardness-contour}. In hard X-rays, the halo gas is not observed because the temperature of the hot halo gas is not high enough. In this case, the brightest parts are close to the disc, where the warm clumps are being lifted. Close to the disc, where the warm clumps are being accelerated by the hot injected gas, the speed of the clumps is small, and the
  relative speed between the clump and the free wind gas is maximum (see equation $4$ in \cite{Sharma2012}) . This makes the bow shock regions close to the disc show up in hard X-rays brighter 
  than the bow shocks far above the disc, where the clumps have reached a larger speed owing to the ram pressure acceleration. 
  
  The third panel of the same figure show the contours of the hardness ratio for the same parameters. We find that the hardest X-rays are radiated by bow shocks
  close to the disc and the top part of the outer shock, owing to the shock heated gas in this region. The right-most panel of this figure compares the hardness ratio
  map with that in CI case. In the central injection case, the hardest X-rays come from the very central region (likely to be absorbed by the ISM) and the shocked
  gas in the shell. In contrast, the DI case shows a hard X-ray spot in along the pole.

  We also show the ratio of the total X-ray luminosity in the $2\hbox{--}5$ keV to that in $0.5\hbox{--}2$ keV range in the left panel of Figure \ref{fig:totLx_soft_hard}.
  The X-ray emission in the CI case is harder than the DI case, as is expected from the temperature ($\ge 10^7$ K) in the central region.  We also see that
  the the hardness in the DI case first increases and then flattens off. Had the only source of X-ray emission been the shocked ISM, then the hardness would 
  have remained a constant, owing to the near constancy of the outer shock speed. However, a significant contribution of X-ray emission comes from the bow shocks
  embedded in the free wind region. We notice that for $L42$ at $\ge 7$ Myr, the hardness increases. This is also the epoch when the gas shocked by the outer shock leaves the simulation box. The increase in the hardness after this epoch implies that the X-ray emitted by the bow shock regions is harder than that
  from the outer shock region, as is also shown by the hardness ratio contour in the right panel of Figure \ref{fig:hardness-contour}. 
  
  In the right panel of Figure \ref{fig:totLx_soft_hard} we show the same ratio as in the left panel, but after subtracting the contribution of the central region. This decreases the 
  hardness of the CI case, as expected. Also, we find that the relative hardness of the DI to the CI case increases with time. This is partly due to the
  fact that bow shock regions in DI cases emit relatively hard X-rays (which are absent in the CI case) and also the outer shock region becomes relatively harder
  for the DI case for the $L42$ case, as predicted by equation \ref{eq:ratio-speed}.
  
\begin{figure*}
	\centering  
	\includegraphics[trim={0 0 0 2cm}, clip=true, width=\linewidth]{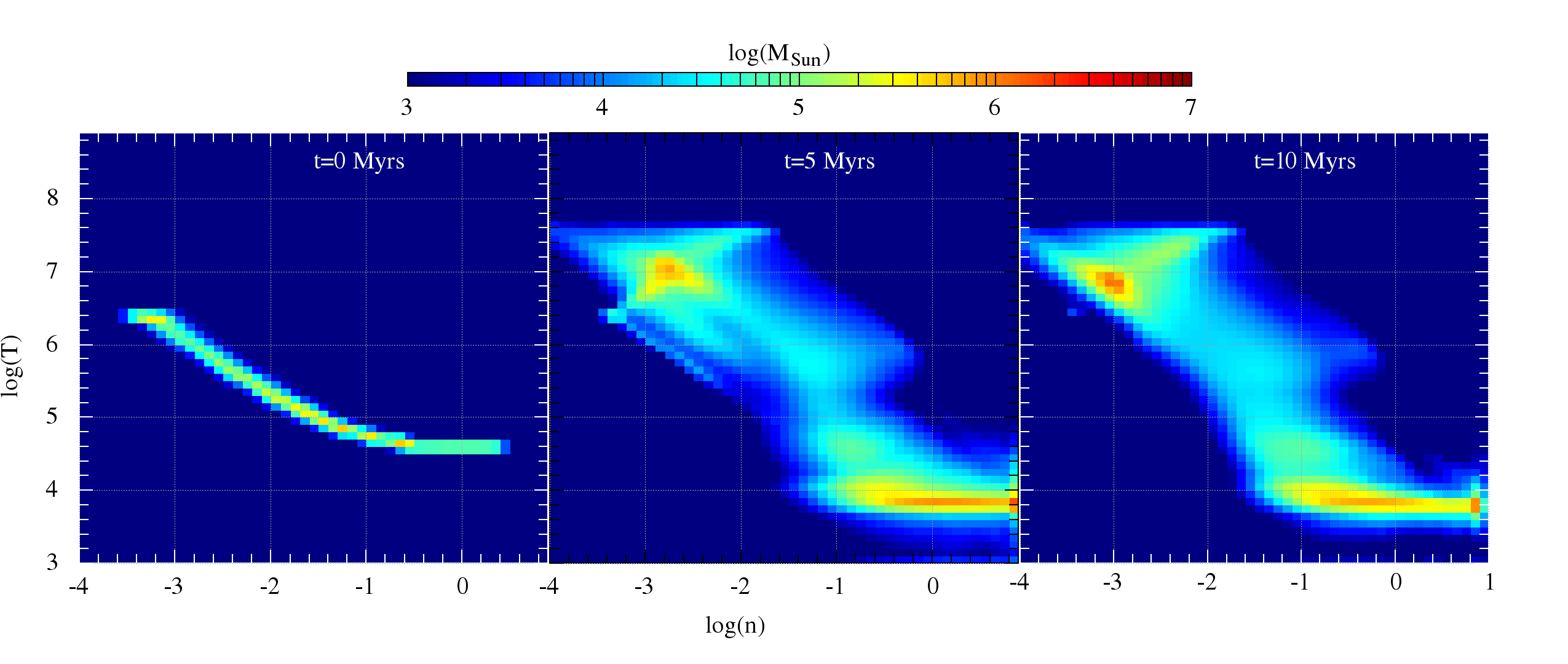}
	\caption{Mass Maps of $L42$-DI at $0$, $5$ and $10$ Myr in the $n-T$ plane. The two peaks at later times represent the hot wind plus bow shocks ($\sim 10^7$ K) and warm clumps ($\sim 10^4$ K)} 
	\label{massmap}
\end{figure*}

  \subsection{Mass of gas at different density and temperature}
  \label{sec:mass-temp}
  It is also important to know at which different temperatures does the mass exist as it directly indicates the wave band of observations aimed to detect this mass. We show the total mass of gas at different densities and temperatures for $L42$ in the case of DI in Figure \ref{massmap}, for $0$, $5$ and $10$ Myr. %Upon comparison with figure 4, which shows distribution of mass with temperature and the number density, we realise that substantial mass does not contribute to X-ray luminosity. The bright yellow patches in the bottom right of the maps at 5 and 10 Myr indicate that significant amount of the gas is cold and dense.%
  The left-most panel shows the initial set up of the gas. The lowest temperature gas (at slightly more than $10^4$ K since it includes the turbulence) resides in the disc and it is also the densest. The highest temperature and the most tenuous gas is located in the halo. The intermediate region in the density-temperature space is occupied by gas in the
  interface region between disc and halo, and this gas is in rough pressure equilibrium, shown by the approximate constancy of the product of density and temperature.  
  
  The right panels of this figure shows that as time progresses, mass shifts to colder and dense regions, which is associated with the formation of the clumps by thermal cooling. In other words, a substantial fraction of the mass in the outflowing gas is located in the warm clumps and in the hot phase generated by the bow shocks (the red regions with $T \sim 10^7$ K). The bright yellow patches on the top left of the maps represent the hot injected material that is following an adiabatic expansion i.e. $T \propto n^{2/3}$ (for $\gamma=5/3$). The intermediate regions ($\sim 10^4-10^6$ K) mostly represent the interface between the cold clumps and the hot wind.
  %These figures show that most of the X-Ray luminosity contribution comes from the disc gas. A parcel is taken to be mostly of disc gas if the disc gas tracer is more than 7.0 and the injected gas tracer is less than 3.0.

  \begin{figure}
   \includegraphics[width=8cm]{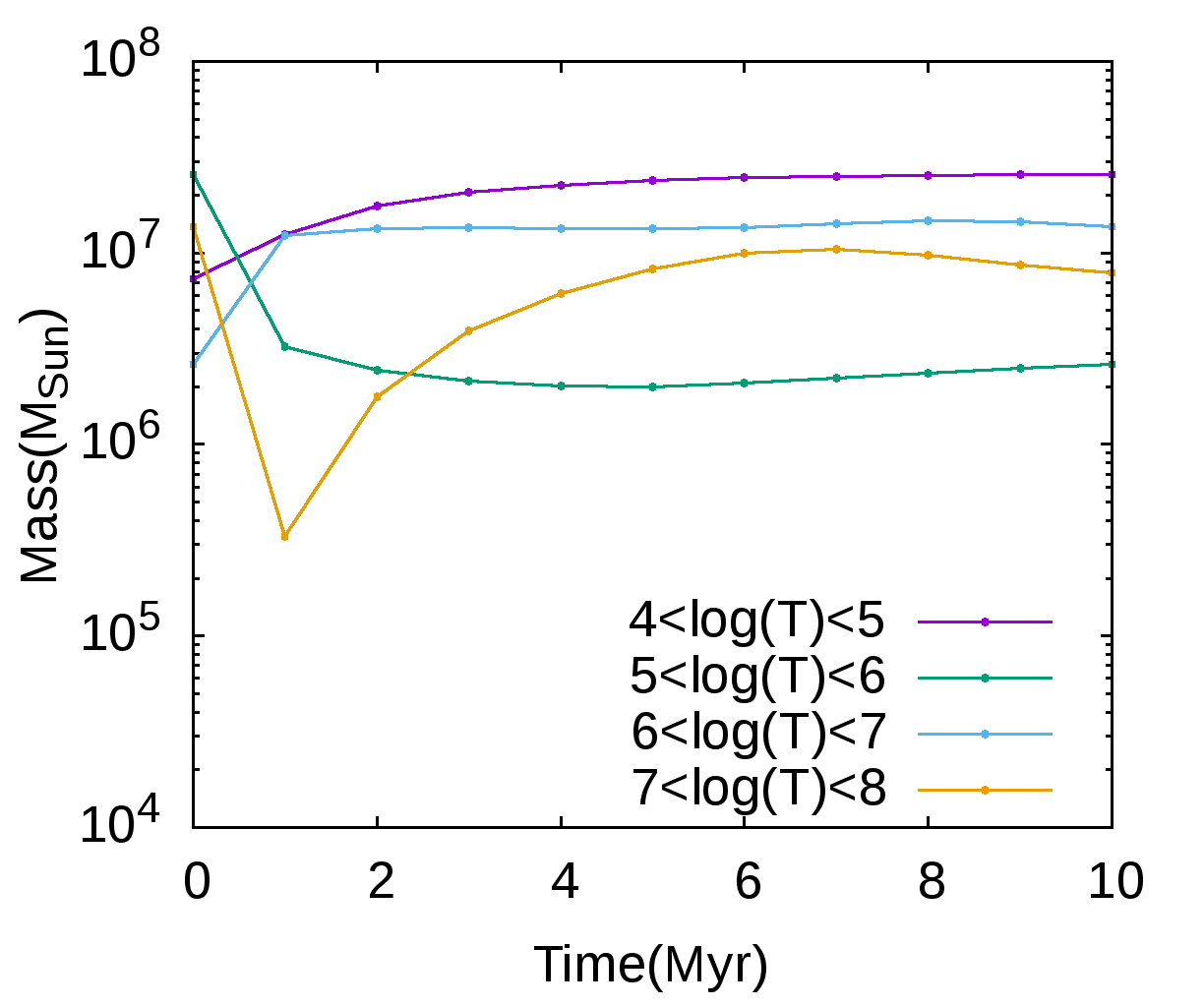}
  \caption{Mass contained in gas in different temperature bins is shown as a function of time for $L42$, DI case ($z>100$ pc).
 }
 \label{fig:mass-temp-bin}
  \end{figure}
  
For an easier comparison of the mass at different phases, we have divided the mass in four temperature bins as shown in Figure \ref{fig:mass-temp-bin}. We find that the mass contained in gas with temperature between $10^6\hbox{--}10^7$ K (responsible for most of the soft X-ray emission) is comparable to the mass in warm clumps ( $\sim 10^4\hbox{--}10^5$ K), and the mass budget remains constant after a few Myr of the onset of star formation. In comparison, the mass contained in gas with temperature in the range of $10^5\hbox{--}10^6$ K is an order of magnitude lower.  In other words, the mass distribution is bimodal in temperature, with peaks at $10^4\hbox{--}10^5$ K and $10^6\hbox{--}10^7$ K suitable for UV and X-ray studies, respectively (see also \cite{KCSI}). 

\begin{figure}
	\centering
	\includegraphics[width=7cm]{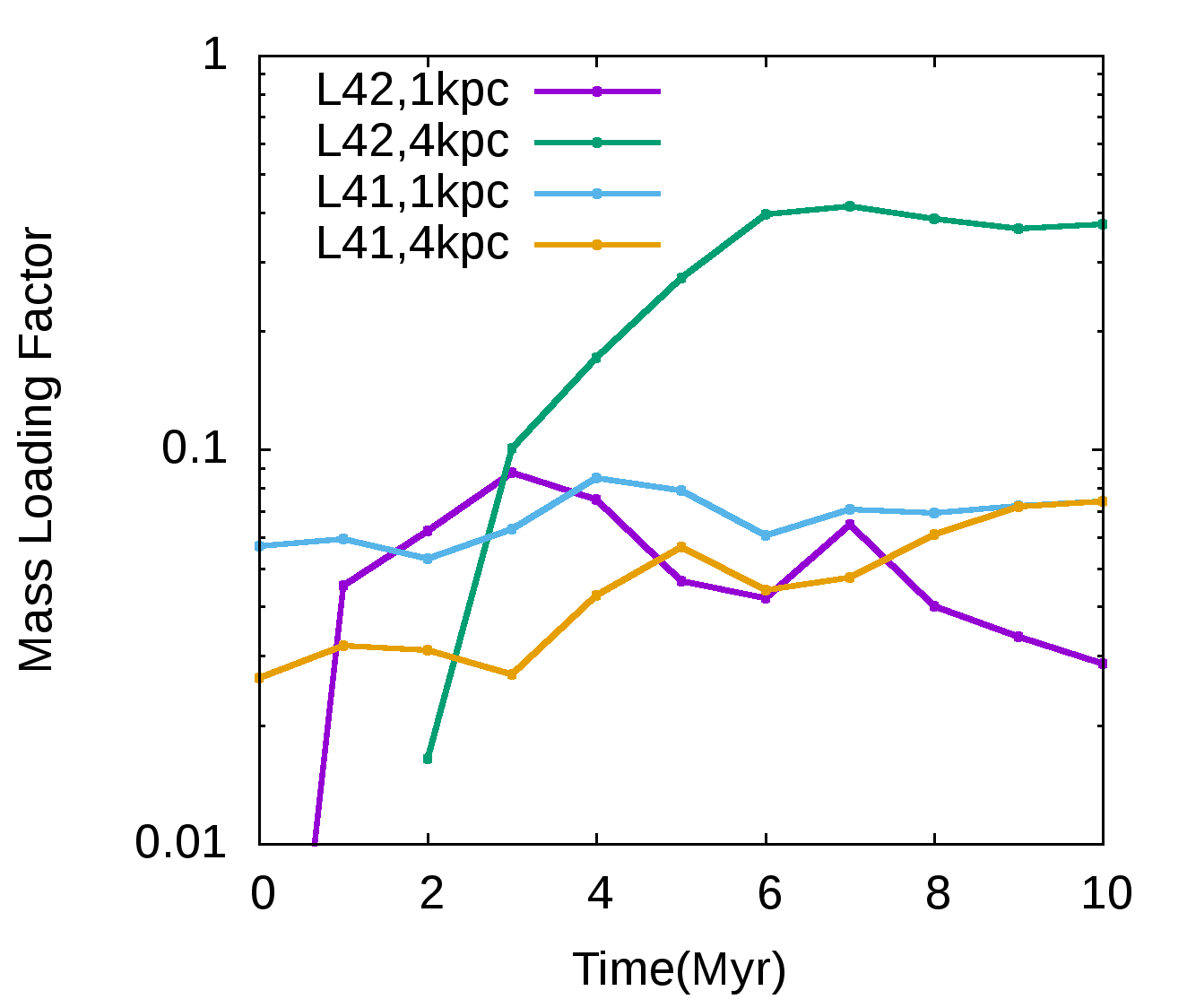}
	\caption{Mass loading factor ($=\dot{M}_{\rm{out}}/\rm{SFR}$) for $L42$ and $L41$ (both DI) for mass flux out of $1$ and $4$ kpc boxes.
	}
	\label{fig:mass-loading}
\end{figure}

\section{Discussion}

\subsection{Mass loading factor}
\label{sec:mlf}
Our main focus in this paper has been on the extra-planar X-ray emission properties of outflows from disc galaxies. In order to study this in an idealized  set-up, we have not followed the evolution of the outflows over a long time scale. This would have entailed the possibility of a partial fall back of outflowing material on the disc, which would further require the coupling of star formation process with gas density, as has been studied in detail by \cite{Fielding2017}. The main focus of these authors has been the dynamical aspect of outflows, and in particular, the relation between the mass loading of outflows and parameters such as surface density of gas. Although this is not our main focus, we have estimated the mass loading factor in our simulation runs, by determining the mass flux out of boxes of different length scales ($1$ and $4$ kpc) as a function of time, which we show in Figure \ref{fig:mass-loading}. It should be borne in mind that the rectangular boxes chosen for mass outflows calculations while outflows themselves follow a near-spherical geometry.

For the $L42$ case, the mass loading factor increases (as the shock sweeps up the halo gas) till the shock fills the $4$ kpc box and thereafter remains at a near constant value for the duration of the run. The difference in the mass loading of $L42$ at $1$ and $4$ kpc is the consequence of entrainment of the disk gas. The mass loading factor for the $L41$ case out of the $4$ kpc box is small because of the fact that the outer shock does not reach this box within the simulation time scale. However there are two important results that we would like to point out here. Firstly, the mass loading factor is of order of few $\times 0.1$, consistent with the results of \cite{Fielding2017} (their figure 3) for a similar time period. Their results show that mass loading increases further if one evolves the outflows for a much longer time period. Secondly, the mass loading factor increases with decreasing SFR (by comparing L41,1 kpc and L42,1kpc). This is also consistent with the trend found by \cite{Heckman2015} (their Figure 6), \cite{Fielding2017}, and predicted by \cite{Sharma2012} (their equation 14). This trend was also noticed in \cite{KCSI} although for centralised injection case.

It should be also noted, as suggested by the discussion in section \ref{sec:mass-temp}, that the total mass loading may be divided almost equally into two phases ($\sim 10^4-10^5$K and $\gtrsim 10^6$ K) and will require multiwavelength observations to reveal the true mass loading in the observed galaxies.

\begin{figure}
	\centering
	\includegraphics[trim={0 0 0 4cm}, clip=true, width=\linewidth]{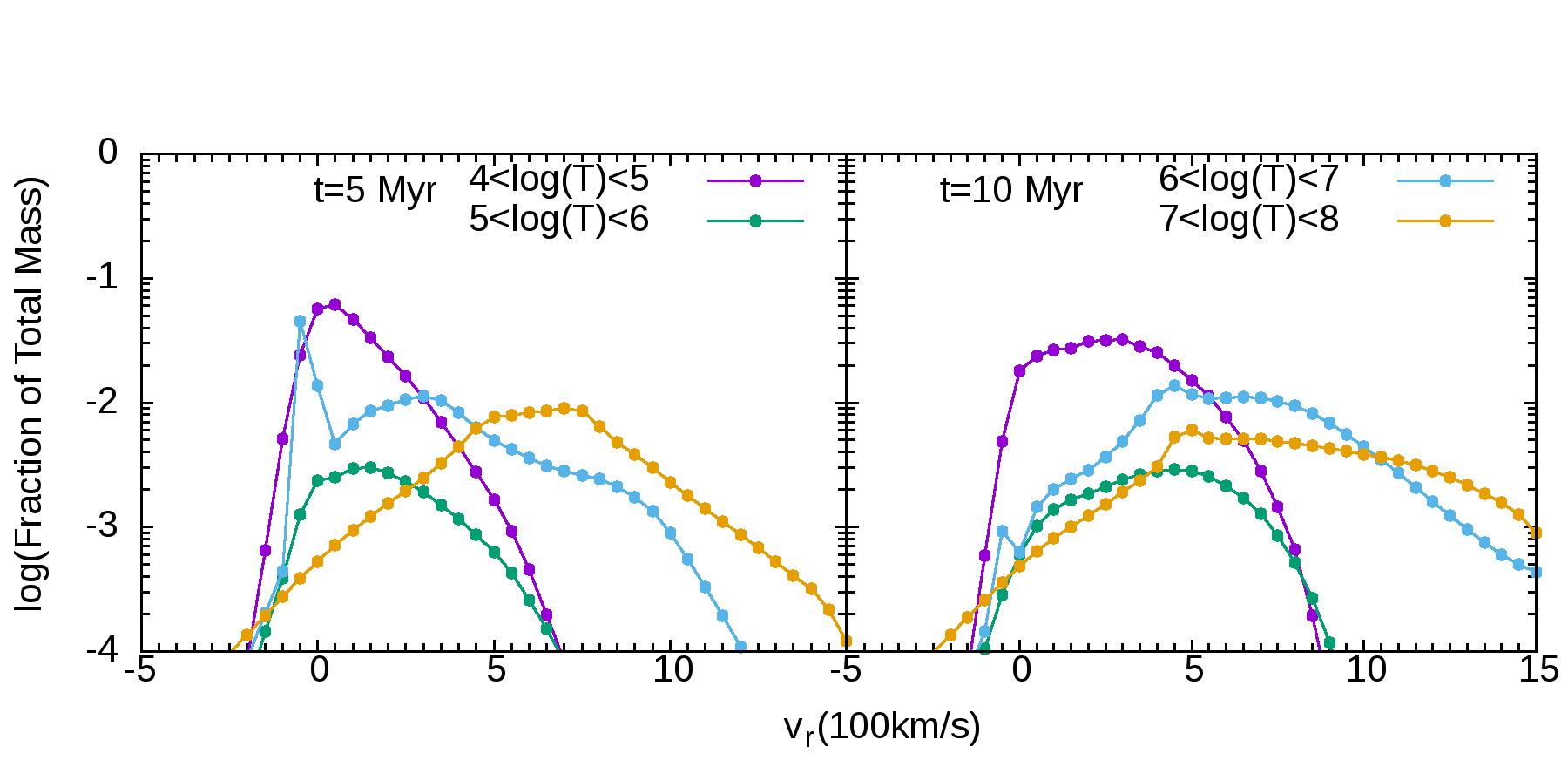}
	\caption{Fraction of total mass contained in a radial speed ($v_r=\vec{v}\cdot \hat{r}$) and temperature bin for $L42$, DI case, for $5$(left) and $10$(right) Myr. The contribution from the disc($z>100$ pc) has been discounted.}
	\label{fig:vel-mass-bin}
\end{figure}

\subsection{Velocity distribution}
Kinematics of the outflowing gas is another parameter that determines if the matter is going to be reaccreted or thrown away far from the centre which can then take part in enriching the CGM with metals and entropy. The dynamics of the warm gas is specifically of interest as it can provide an estimate of how much of this gas will `rain back' to the stellar disc and participate in further star formation. 
	
Fig. \ref{fig:vel-mass-bin} illustrates the fraction of mass contained in different radial speed ($\rm{v}_r=\vec{\rm{v}}\cdot \hat{r}$) and temperature bins for gas above $z = 100$ pc. 
%\sout{, is shown in figure \ref{fig:vel-mass-bin}. Cold gas is slower while hot gas is faster. As time progresses, the cold clumps are accelerated by the hot gas and occupy a wider range in speed.} 
We notice that the warm clumps ($\sim 10^4-10^5$K) have a very narrow velocity range that extends from $\sim -100$ (fountain flow) to $\sim 400$ \kmps. The range, however, increases at a later time and reaches $\sim 800$ \kmps at $10$ Myr. This can be attributed to the entrainment of the clumps by a fast moving ($\sim 1000$ \kmps) hot gas ($\gtrsim 10^6$ K). The $> 10^6$ K gas represents the hot wind, bow shocks and also the CGM. The peak at $\sim 0$ \kmps for the $10^6-10^7$K gas at $5$ Myr is due to the stationary CGM which has not yet been shocked yet. This peak disappears at a later time indicating that the shock has crossed the boundary of the computational box. A large outflowing velocity of the hot gas means that this gas can escape the gravitational potential of the galaxy ($\sim 300$ \kmps) and mix with the CGM and increase the CGM metallicity and entropy. 

Note that this picture of a multiphase wind is very different from a single fluid approximation of the steady wind \citep{chevalier1985, Strickland2009, Zhang2014} where the temperature and density are functions of radius alone. This model therefore predicts a single velocity at a distance which depends on the thermalisation efficiency and the mass loading factor of the wind. The multiphase wind model, on the other hand, is fundamentally different as there can be multiple velocities depending on the phase coexisting at any radius. This naturally predicts that a certain fraction of the lower velocity warm clumps is likely to be reaccreted but the hot and tenuous gas is most likely to be thrown away into the CGM.

We also find a good fraction of the warm clumps to have an outflowing velocity $\gtrsim$ the escape velocity ($\sim 300$ \kmps), meaning that these clumps can escape the galaxy and enrich the IGM. However, the real scenario can be different. As these clumps rise in height, the ram pressure of the wind falls too rapidly to accelerate the clumps further. From there onwards, the clumps follow almost a ballistic trajectory and travel outwards through the hot CGM. It has been shown in several numerical simulations that such a journey causes Kelvin-Helmholtz instability and conductive evaporation of the clump \citep{Cooper2009, Armillotta2017}. It is, however, uncertain if the results will remain same in the presence of the magnetic field. 

\subsection{Testing a multiphase wind model in observation}
It is clear from the observations of the co-existence between H$\alpha$ and soft X-ray emission \citep{Grimes2005} that a multiphase model, like the one presented here, should be considered while studying the extra-planar X-ray emission. This model, however, can be further tested independently by considering the hardness ratio of the extra-planar emission. As shown in Fig. \ref{fig:totLx_soft_hard}, that the hardness ratio of the extra-planar emission ($|z| \gtrapprox 100$ pc, right panel) for L42-DI case is much larger than a L42-CI case. This is due to the presence of the bow shocks. We propose that a distinction between the classical wind model versus a multiphase model can be done in such an observation.

\section{Conclusions}
In summary, we present 3D numerical simulations to describe a multiphase nature of the extra-planar wind and related X-ray emission. Our main findings are:

\begin{enumerate}
\item Distribution of energy and mass sources throughout the disc of a galaxy significantly changes the morphology and structure of outflows generated. These changes are a direct consequence of the formation of warm clumps in the free wind region of the outflows. We attribute these clumps to  disc gas which is lifted up by ram pressure. ISM turbulence causes a disc to be naturally inhomogeneous and formation of clumps in such a disc has been studied earlier. Our work, however, shows that the clumps are formed even for a homogeneous and stratified disc. 
 
\item The warm clumps of disc gas once lifted interact with the hot free wind. The relative speed between the two leads to formation of bow shocks.  These are the primary source of soft ($0.5\hbox{--}2.0$ keV) X-rays, since the intermediate temperature region contains most of the mass of the outflow. Such a process is absent in the case where the energy and mass deposition takes place at the centre of the disc (CI). In CI, thus, the strongest contributor to the soft X-rays is the central region itself, followed by the shocked shell. This produces a sharp fall in the brightness profile along the minor axis of the galaxy. However, in the distributed injection case, the bow shock regions in the free wind emit copious X-rays and make the brightness profile to decrease slowly, consistent with observations. This model of multiphase wind is fundamentally different from single fluid models and has the potential to better explain the observed extra-planar X-ray emission
 
\item We find that the bow shock regions also emit hard X-rays ($2\hbox{--}5$ keV), particularly those close to the disc. 
By assuming plane parallel symmetry we can estimate the speed of the shock in both cases. We find that while the shock speed decreases with time for CI, it remains roughly a constant for DI. This is also borne by the hardening of the X-rays for  DI as time progresses. The hardness ratio is  another handle that can be used to distinguish between a single fluid model of galactic wind vs a multiphase nature of the wind.
\end{enumerate}

Based on the current results, we believe that a detailed study of such multiphase wind models incorporating self-consistent star formation and analysis of the mock X-ray spectrum will lead us towards a better understanding of the extra-planar X-ray emission and unlock several mysteries related to galactic feedback and CGM enrichment.

\section*{Acknowlegdments}
We would like to thank Yuval Birnboim, Siddhartha Gupta and Saurabh Singh for many useful suggestions and discussions. The simulations were performed on the SaharaT cluster at the Supercomputer Education and Research Centre, Indian Institute of Science. KCS thanks the hospitality of the Kavli Institute of Theoretical Physics, Santa Barbara and acknowledges a partial support by the National Science Foundation under Grant No. NSF PHY-1125915. KCS is also partly supported by the Israeli Centers of Excellence (I-CORE) 
program (center no. 1829/12) and by the Israeli Science Foundation (ISF grant no. 857/14). PS acknowledges an India-Israel joint research grant (6-10/2014[IC]). YS acknowledges support from RFBR (project code 17-52-45053).
The authors would like to thank the referee, Geoffrey V. Bicknell, for his valuable comments.

%\section{Appendix}

%The above analysis tells us that for DI, $\dot{R}_{DI} \propto \Sigma_L ^\frac{1}{3}$, which gives a reduction in speed by a factor of $2$, between L42($1000$\kmps) and L41($360$\kmps).
%%%%%%%%%%%%%%%%%%%%%%%%%%%%%%%%%%%%%%%%%%%%%%%%%%



%\bibliographystyle{mnras}

\begin{thebibliography}{99}
	
\bibitem[\protect\citeauthoryear{Aguirre et al.}{2001}]{aguirre2001}
Aguirre, A., Hernquist, L., Schaye, J., Katz, N., Weinberg, D. H., Gardner, 2001,
\ufhref[webgreen]{http://adsabs.harvard.edu/abs/2001ApJ...561..521A}{ApJ}, 
\ufhref[webgreen]{http://dx.doi.org/10.1086/323370}{561,521}

\bibitem[\protect\citeauthoryear{Armillotta \etal}{2017}]{Armillotta2017}
Armillotta L., Fraternali F., Werk J. K., Prochaska J. X., Marinacci F., 2017,
\ufhref[webgreen]{http://adsabs.harvard.edu/abs/2017MNRAS.470..114A}{MNRAS, 470, 114}


%\bibitem[\protect\citeauthoryear{Bett \etal}{2010}]{Bett2010}
%Bett P., Eke V., Frenk C. S., Jenkins A., Okamoto T., 2010
%\ufhref[webgreen]{https://doi.org/10.1111/j.1365-2966.2010.16368.x}{MNRAS, 404, 1137}

\bibitem[\protect\citeauthoryear{Bigiel et al.}{2008}]{bigiel2008}
Bigiel, F.; Leroy, A.; Walter, F.; Brinks, E.; de Blok, W. J. G.; Madore, B.; Thornley, M. D., 2008,
\ufhref[webgreen]{http://adsabs.harvard.edu/abs/2008AJ....136.2846B}{ApJ}, 
\ufhref[webgreen]{http://iopscience.iop.org/article/10.1088/0004-6256/136/6/2846/meta}{136, 2846}

\bibitem[\protect\citeauthoryear{Bustard \etal}{2016}]{Bustard2016}
Bustard C., Zweibel E. G., D' Onghia E., 2016, ApJ, 819, 29
\ufhref[webgreen]{http://adsabs.harvard.edu/abs/2016ApJ...819...29B}{ApJ}, 
\ufhref[webgreen]{http://iopscience.iop.org/article/10.3847/0004-637X/819/1/29/pdf}{819, 29}

\bibitem[\protect\citeauthoryear{Chevalier \& Clegg}{1985}]{chevalier1985}
Chevalier, R. A., Clegg, A. W., 1985, 
\ufhref[webgreen]{https://www.nature.com/nature/journal/v317/n6032/abs/317044a0.html}{Nature, 317, 44}

%\bibitem[\protect\citeauthoryear{Clarke \& Oey}{2002}]{Clarke2002}
%Clarke, C., Oey, S. 2002, 
%\ufhref[webgreen]{http://adsabs.harvard.edu/abs/2002MNRAS.337.1299C}{MNRAS},
%\ufhref[webgreen]{https://academic.oup.com/mnras/article-lookup/doi/10.1046/j.1365-8711.2002.05976.x}{337, 1299}


\bibitem[\protect\citeauthoryear{Cooper et al.}{2008}]{Cooper2008}
Cooper, J. L., Bicknell, G. V., Sutherland, R. S., Bland-Hawthorn, J. 2008, 
\ufhref[webgreen]{http://adsabs.harvard.edu/abs/2008ApJ...674..157C}{ApJ},
\ufhref[webgreen]{http://iopscience.iop.org/article/10.1086/524918/meta}{674, 157}

\bibitem[\protect\citeauthoryear{Cooper et al.}{2009}]{Cooper2009}
Cooper, J. L., Bicknell, G. V., Sutherland, R. S., Bland-Hawthorn,
\ufhref[webgreen]{http://adsabs.harvard.edu/abs/2009ApJ...703..330C}{ApJ},
\ufhref[webgreen]{http://iopscience.iop.org/article/10.1088/0004-637X/703/1/330/meta}{703, 330}, 


\bibitem[\protect\citeauthoryear{Dekel \& Silk}{1986}]{Dekel1986}
Dekel A., Silk, J. 1986,
\ufhref[webgreen]{http://adsabs.harvard.edu/abs/1986ApJ...303...39D}{ApJ},
\ufhref[webgreen]{http://articles.adsabs.harvard.edu/cgi-bin/nph-iarticle_query?1986ApJ...303...39D&amp;data_type=PDF_HIGH&amp;whole_paper=YES&amp;type=PRINTER&amp;filetype=.pdf}{303, 39}


\bibitem[\protect\citeauthoryear{Ferrara et al.}{2000}]{ferrara2000}
Ferrara, A., Pettini, M., Shchekinov, Y., 2000
\ufhref[webgreen]{http://adsabs.harvard.edu/abs/2000MNRAS.319..539F}{MNRAS},
\ufhref[webgreen]{http://iopscience.iop.org/article/10.1046/j.1365-8711.2000.03857.x}{319, 539} 


\bibitem[\protect\citeauthoryear{Fielding et al.}{2017}]{Fielding2017}
Fielding, D., Quataert, E, Martizzi, D., Faucher-Gigu\`ere, C. 2017, 
\ufhref[webgreen]{http://adsabs.harvard.edu/abs/2017MNRAS.470L..39F}{MNRAS}, 
\ufhref[webgreen]{https://academic.oup.com/mnrasl/article-lookup/doi/10.1093/mnrasl/slx072}{470, L39}

\bibitem[\protect\citeauthoryear{Guo et al.}{2012}]{Guo2012}
Guo, F., Mathews, W., Dobler G., Oh P., 2012
\ufhref[webgreen]{http://adsabs.harvard.edu/abs/2012ApJ...756..182G}{ApJ}, 
\ufhref[webgreen]{http://iopscience.iop.org/article/10.1088/0004-637X/756/2/182/meta}{756, 2}

\bibitem[\protect\citeauthoryear{Grimes \etal}{2005}]{Grimes2005}
Grimes J. P., Heckman T., Strickland D., Ptak A., 2005, ApJ, 628, 187 
\ufhref[webgreen]{http://adsabs.harvard.edu/abs/2005ApJ...628..187G}{ApJ}, 
\ufhref[webgreen]{http://iopscience.iop.org/article/10.1086/430692/pdf}{756, 2}

\bibitem[\protect\citeauthoryear{Heald \etal}{2007}]{Heald2007}
Heald G. H., Rand R., Benjamin R. A., Bershady M. A., 2007, 
\ufhref[webgreen]{http://adsabs.harvard.edu/abs/2007ApJ...663..933H}{ApJ}, 
\ufhref[webgreen]{http://iopscience.iop.org/article/10.1086/518087/meta}{663, 933}

\bibitem[\protect\citeauthoryear{Heckman et al.}{2015}]{Heckman2015}
Heckman, T. M., Alexandroff, R. M., Borthakur, S., Overzier, R., Leitherer, C. 2015, 
\ufhref[webgreen]{http://adsabs.harvard.edu/abs/2015ApJ...809..147H}{ApJ},
\ufhref[webgreen]{http://iopscience.iop.org/article/10.1088/0004-637X/809/2/147/meta}{809, 147}



\bibitem[\protect\citeauthoryear{Higdon \& Lingenfelter}{2013}]{higdon2013}
Higdon, J. C., Lingenfelter, R. E. 2013, 
\ufhref[webgreen]{http://adsabs.harvard.edu/abs/2013ApJ...775..110H}{ApJ},
\ufhref[webgreen]{http://iopscience.iop.org/article/10.1088/0004-637X/775/2/110/meta}{ 775, 110}


\bibitem[\protect\citeauthoryear{Kalberla \& Kerp}{2009}]{Kalberla2009}
Kalberla M. W., Kerp J., 2009, ARAA, 47, 27
\ufhref[webgreen]{http://adsabs.harvard.edu/abs/2009ARA/26A..47...27K}{ARAA},
\ufhref[webgreen]{http://www.annualreviews.org/doi/full/10.1146/annurev-astro-082708-101823}{47, 27}


\bibitem[\protect\citeauthoryear{Kennicutt}{1998}]{kennicutt1998}
Kennicutt, Jr. R. C. 1998, 
\ufhref[webgreen]{http://adsabs.harvard.edu/abs/1998ApJ...498..541K}{ApJ},
\ufhref[webgreen]{http://iopscience.iop.org/article/10.1086/305588/meta	}{498, 541}


\bibitem[\protect\citeauthoryear{Larson}{1974}]{larson1974}
Larson, R. 1974, 
\ufhref[webgreen]{http://adsabs.harvard.edu/abs/1974MNRAS.169..229L}{MNRAS},
\ufhref[webgreen]{https://academic.oup.com/mnras/article-lookup/doi/10.1093/mnras/169.2.229}{4169, 229}



\bibitem[\protect\citeauthoryear{Li \& Wang}{2013}]{Li2013}
Li J., Wang Q. D. 2013, 
\ufhref[webgreen]{https://academic.oup.com/mnras/article/428/3/2085/1065231/Chandra-survey-of-nearby-highly-inclined-disc}{MNRAS, 428, 2085}


\bibitem[\protect\citeauthoryear{Mignone et al.}{2007}]{Mignone2007}
Mignone A., Bodo G., Massaglia S., Matsakos T., Tesileanu O., Zanni C., Ferrari A., 2007, 
\ufhref[webgreen]{http://adsabs.harvard.edu/abs/2007ApJS..170..228M}{ApJSS},
\ufhref[webgreen]{http://iopscience.iop.org/article/10.1086/513316/meta	}{ 170, 228}


\bibitem[\protect\citeauthoryear{Miyamoto \& Nagai}{1975}]{Miyamoto1975}
Miyamoto M., Nagai R. 1975, 
\ufhref[webgreen]{http://adsabs.harvard.edu/abs/1975PASJ...27..533M}{PASJ},
\ufhref[webgreen]{http://adsabs.harvard.edu/abs/1975PASJ...27..533M}{27, 553}

\bibitem[\protect\citeauthoryear{Nath \& Trentham}{1997}]{nath1997}
Nath B. B., Trentham, N. 1997, 
\ufhref[webgreen]{http://adsabs.harvard.edu/abs/1997MNRAS.291..505N}{MNRAS},
\ufhref[webgreen]{https://academic.oup.com/mnras/article-lookup/doi/10.1093/mnras/291.3.505}{	 291, 505}


\bibitem[\protect\citeauthoryear{Nath \& Shchekinov}{2013}]{nath2013}
Nath B. B., Shchekinov Y., 2013, 
\ufhref[webgreen]{http://adsabs.harvard.edu/abs/2013ApJ...777L..12N}{ApJ},
\ufhref[webgreen]{http://iopscience.iop.org/article/10.1088/2041-8205/777/1/L12/meta}{777, L12}

\bibitem[\protect\citeauthoryear{Navarro et al}{1997}]{Navarro1997}
Navarro J. F., Frenk C. S., White, D. M. 1997, 
\ufhref[webgreen]{http://adsabs.harvard.edu/abs/1997ApJ...490..493N}{ApJ},
\ufhref[webgreen]{http://iopscience.iop.org/article/10.1086/304888/meta	}{490, 493}



\bibitem[\protect\citeauthoryear{Oppenheimer \& Dav\'e}{2006}]{Oppenheimer2006}
Oppenheimer, B. D., Dav\'e, R. 2006, 
\ufhref[webgreen]{http://adsabs.harvard.edu/abs/2006MNRAS.373.1265O}{MNRAS},
\ufhref[webgreen]{https://academic.oup.com/mnras/article-lookup/doi/10.1111/j.1365-2966.2006.10989.x}{373, 1265}


%\bibitem[\protect\citeauthoryear{Roy et al.}{2015}]{Roy2015}
% Roy, A., Nath, B. B., Sharma, P. 2015, 
%\ufhref[webgreen]{http://adsabs.harvard.edu/abs/2015MNRAS.451.1939R}{MNRAS},
%\ufhref[webgreen]{https://academic.oup.com/mnras/article/451/2/1939/986111/Narrow-escape-how-ionizing-photons-escape-from}{451, 1939}
 
 
\bibitem[\protect\citeauthoryear{Sarkar et al.}{2015}]{KCSI}
Sarkar, K . C., Nath, B. B., Sharma P., Shchekinov Y. 2015, 
 \ufhref[webgreen]{http://adsabs.harvard.edu/abs/2015MNRAS.448..328S}{MNRAS},
\ufhref[webgreen]{https://academic.oup.com/mnras/article/448/1/328/1750379/Long-way-to-go-how-outflows-from-large-galaxies}{448, 328}


\bibitem[\protect\citeauthoryear{Sarkar et al.}{2016}]{Sarkar2016}
Sarkar, K . C., Nath, B. B., Sharma P., Shchekinov Y. 2016, 
 \ufhref[webgreen]{http://adsabs.harvard.edu/abs/2016ApJ...818L..24S}{ApJ},
\ufhref[webgreen]{http://iopscience.iop.org/article/10.3847/2041-8205/818/2/L24/meta}{818, L2}

\bibitem[\protect\citeauthoryear{Sharma \& Nath}{2012}]{Sharma2012}
Sharma, M., Nath, B. B. 2012, 
\ufhref[webgreen]{http://adsabs.harvard.edu/abs/2015MNRAS.448..328S}{ApJ},
\ufhref[webgreen]{http://iopscience.iop.org/article/10.1088/0004-637X/750/1/55/meta}{750, 55}


\bibitem[\protect\citeauthoryear{Sharma et al.}{2014}]{Sharma2014}
Sharma, P., Roy, A., Nath, B. B., Shchekinov, Y. 2014, 
\ufhref[webgreen]{http://adsabs.harvard.edu/abs/2014MNRAS.443.3463S}{MNRAS},
\ufhref[webgreen]{https://academic.oup.com/mnras/article/443/4/3463/1008398/In-a-hot-bubble-why-does-superbubble-feedback-work}{443, 3463}


%\bibitem[\protect\citeauthoryear{Silk}{1997}]{Silk1997}
%Silk, J. 1997, 
%\ufhref[webgreen]{http://adsabs.harvard.edu/abs/1997ApJ...481..703S}{ApJ},
%\ufhref[webgreen]{http://iopscience.iop.org/article/10.1086/304073/meta}{ 481, 703}


\bibitem[\protect\citeauthoryear{Strickland et al.}{2002}]{strickland2002}
Strickland, D. K., Heckman, T. M., Weaver, K. A., Hoopes, C. G., Dahlem, M. 2002, 
\ufhref[webgreen]{http://iopscience.iop.org/article/10.1086/338889/meta}{ApJ}, 
\ufhref[webgreen]{http://adsabs.harvard.edu/abs/2002ApJ...568..689S}{481, 703}

\bibitem[\protect\citeauthoryear{Strickland \etal}{2004}]{Strickland2004}
{Strickland}, D.~K. and {Heckman}, T.~M. and {Colbert}, E.~J.~M. and 
{Hoopes}, C.~G. and {Weaver}, K.~A., 2002,
\ufhref[webgreen]{http://adsabs.harvard.edu/abs/2004ApJ...606..829S}{ ApJ, 606, 829 }

\bibitem[\protect\citeauthoryear{Strickland \& Heckman}{2007}]{strickland2007}
Strickland, D. K., Heckman, T. M. 2007, 
\ufhref[webgreen]{http://iopscience.iop.org/article/10.1086/338889/meta}{ApJ}, 
\ufhref[webgreen]{http://adsabs.harvard.edu/abs/2007ApJ...658..258S}{658, 258}

\bibitem[\protect\citeauthoryear{Strickland \& Heckman}{2009}]{Strickland2009}
Strickland, D. K., Heckman, T. M. 2009, 
\ufhref[webgreen]{http://labs.adsabs.harvard.edu/ui/abs/2009ApJ...697.2030S}{ApJ, 697, 2030}

\bibitem[\protect\citeauthoryear{Strickland \& Heckman}{2009}]{Strickland2000}
Strickland, D. K., Stevens, I. R. 2000, 
\ufhref[webgreen]{http://adsabs.harvard.edu/abs/2000MNRAS.314..511S}{MNRAS, 697, 2030}
\ufhref[webgreen]{https://arxiv.org/pdf/astro-ph/0001395.pdf}{314, 511S}

\bibitem[\protect\citeauthoryear{Sutherland \&  Bicknell}{2007}]{Sutherland2007}
Sutherland, R. S.,Bicknell, G. V. 2007, 
\ufhref[webgreen]{http://iopscience.iop.org/article/10.1086/520640/meta}{ApJ}
\ufhref[webgreen]{http://adsabs.harvard.edu/abs/2007ApJS..173...37S}{173, 37}

\bibitem[\protect\citeauthoryear{Tahara et al.}{2015}]{Tahara2015}
Tahara, M. et al. 2015,
\ufhref[webgreen]{http://adsabs.harvard.edu/abs/2015ApJ...802...91T}{ApJ}, 
\ufhref[webgreen]{http://iopscience.iop.org/article/10.1088/0004-637X/802/2/91/meta}{802, 91}

\bibitem[\protect\citeauthoryear{Thompson et al.}{2016}]{Thompson2016}
Thompson, T. A., Quataert, E., Zhang, D., Weinberg, D. H. 2016, 
\ufhref[webgreen]{http://adsabs.harvard.edu/abs/2016MNRAS.455.1830T}{MNRAS}, 
\ufhref[webgreen]{https://academic.oup.com/mnras/article/455/2/1830/1119214}{455, 1830}

\bibitem[\protect\citeauthoryear{Wang \etal}{2016}]{Wang2016}
Wang Q. D., Li J., Jiang X., Fang T., 2016, MNRAS, 457, 1385
\ufhref[webgreen]{http://adsabs.harvard.edu/abs/2016MNRAS.457.1385W}{MNRAS}, 
\ufhref[webgreen]{https://academic.oup.com/mnras/article/457/2/1385/965244}{457, 1385}

\bibitem[\protect\citeauthoryear{Weaver et al.}{1977}]{Weaver1977}
Weaver, R., McCray, R., Castor, J., Shapiro, P., Moore, R. 1977,
\ufhref[webgreen]{http://adsabs.harvard.edu/abs/1977ApJ...218..377W}{ApJ}, 
\ufhref[webgreen]{http://articles.adsabs.harvard.edu/cgi-bin/nph-iarticle_query?1977ApJ...218..377W&amp;data_type=PDF_HIGH&amp;whole_paper=YES&amp;type=PRINTER&amp;filetype=.pdf}{218, 377}



\bibitem[\protect\citeauthoryear{Zhang et al.}{2014}]{Zhang2014}
Zhang, D., Thompson, T. A., Murray, N., Quataert, E. 2014,
\ufhref[webgreen]{http://adsabs.harvard.edu/abs/2014ApJ...784...93Z}{ApJ}, 
\ufhref[webgreen]{http://iopscience.iop.org/article/10.1088/0004-637X/784/2/93/meta}{784, 93}

\end{thebibliography}


%%%%%%%%%%%%%%%%%%%%%%%%%%%%%%%%%%%%%%%%%%%%%%%%%%

%%%%%%%%%%%%%%%%% APPENDICES %%%%%%%%%%%%%%%%%%%%%

\appendix

\section{Distribution of Points}
\label{injection-points}
\begin{figure*}
 \minipage{0.24\textwidth}
\includegraphics[width=\textwidth]{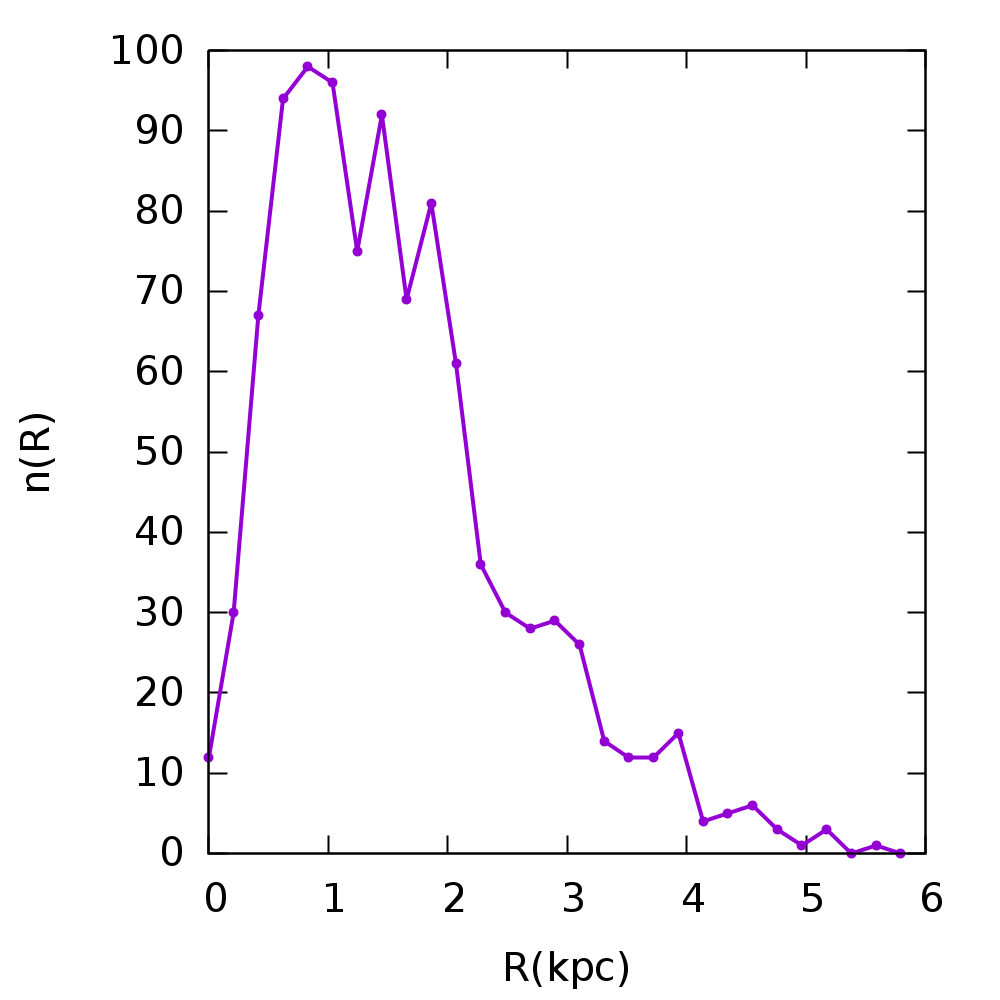}
\endminipage
\minipage{0.24\textwidth}
\includegraphics[width=\textwidth]{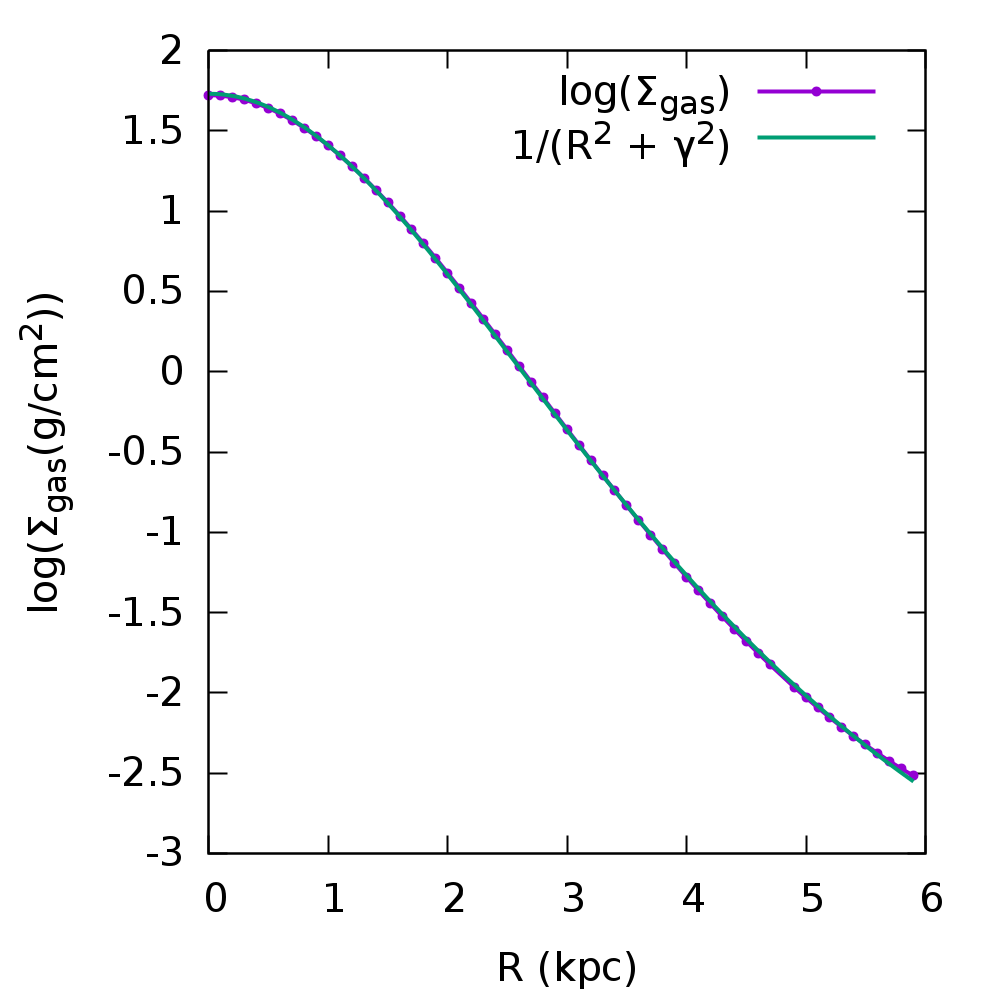}
\endminipage
 \caption{From left to right-(a)Number of points distributed along the R axis;(b)Surface Density of gas plotted against the distance along the R axis.} 
\label{fig:nRsigR}
 \end{figure*}
The Kennicutt-Schmidt law($\Sigma_{\rm SFR} \propto \Sigma_{\rm gas}^{1.4}$) is putatively a relation between the star formation in the disc of a galaxy and the star formation rate in the disc. We integrated the gas density of the disc along the $z$ direction. The curve, thus found, fits to a Lorentzian against the cylindrical distance on a semi-log scale, as shown in Figure \ref{fig:nRsigR}, left panel. We find that the $\Sigma_{\rm gas}$ is related to $R$ in the following manner,
\be
{\rm log}(\Sigma_{\rm gas}) \sim {{\gamma^2 \beta} \over{ R^2 + \gamma^2 } }\,,
\ee
where $\gamma$ and $\beta$ are parameters that are found from the fit as $5.5$ kpc and $6.0$ kpc, respectively. From the Kennicutt-Schimdt law we can write that,
\be
{\rm log}(\Sigma_{\rm SFR}) \sim {{1.4\gamma^2 \beta} \over{ R^2 + \gamma^2 } }\,,
\ee

For simplicity, we have assumed that all the 1000 injection points have identical energy input, $\Sigma_{\rm SFR}$ can be written in terms of the number density of injection points, $n(R)$, in the following manner,
\be
\Sigma_{\rm SFR} = \frac{n(R) dR}{2 \pi R dR} \,.
\ee

The expression for n($R$) then becomes,
\be
%\begin{center}
n(R) = A R \exp \left[\frac{1.4 \gamma^2 \beta}{R^2 + \gamma^2}\right] \,.
%\end{center}
\label{nR}
\ee
Here the parameters $\gamma$ and $\beta$ are found using a fitting function. For our case the values of $\gamma$ and $\beta$ are $5.5$ kpc and $6.0$ kpc, respectively. The parameter $A$ can be found using the normalization,  
\be
\int_0^6 n(R)dR = 1000= \int_0^6 A R \exp \big[\frac{1.4 \gamma^2 \beta}{R^2 + \gamma^2}\big] dR \,.
\ee
It should be noted here that n($R$) is not monotonic and exhibits a maximum. For the parameters we have chosen, the maxima is $\sim 1.0$ kpc. Along the $x$ and $y$ direction these points have a uniform random distribution. Along the $z$-axis, the number density of points has been pegged to the underlying gas density along the same axis. Thus, the probability of finding an injection at a height z, P(z), above the mid-plane($z=0$) is given by-
\be
P(z) = \frac{\rho(z)}{\rho(z=0)}
\ee

\begin{figure*}
\includegraphics[width=\linewidth]{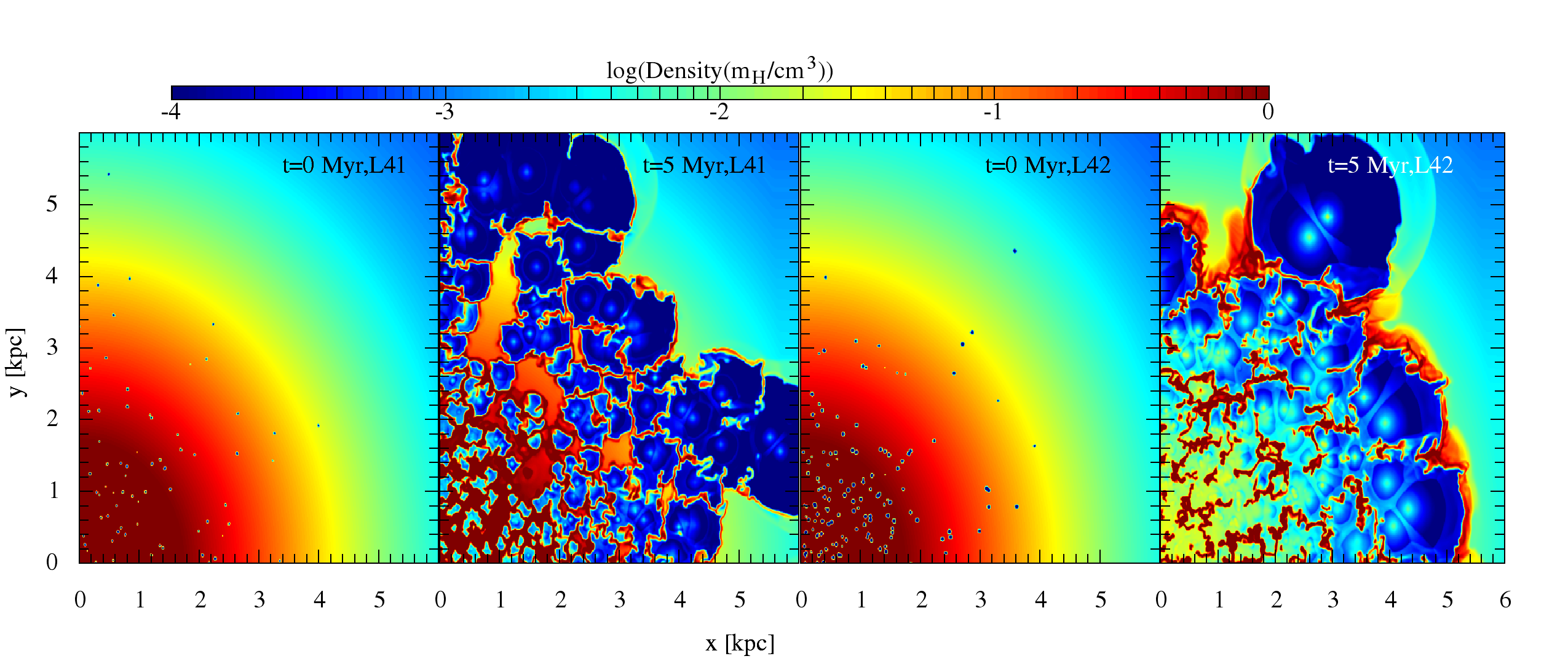}
 \caption{From left to right-(a)The density distribution in the $z=0$ plane at $t=0$ Myr for $L42$, DI;(b) Same as (a), but for $5$ Myr.} 
\label{fig:xyplane}
 \end{figure*}

Snapshots of density distribution in the $x-y$ plane at $t=0,5$ Myr are shown in Figure \ref{fig:xyplane} for L41 and L42 cases.
\section{Resolution Study}
\label{appendix:resolution-study}
\begin{figure*}
\includegraphics[width=\linewidth]{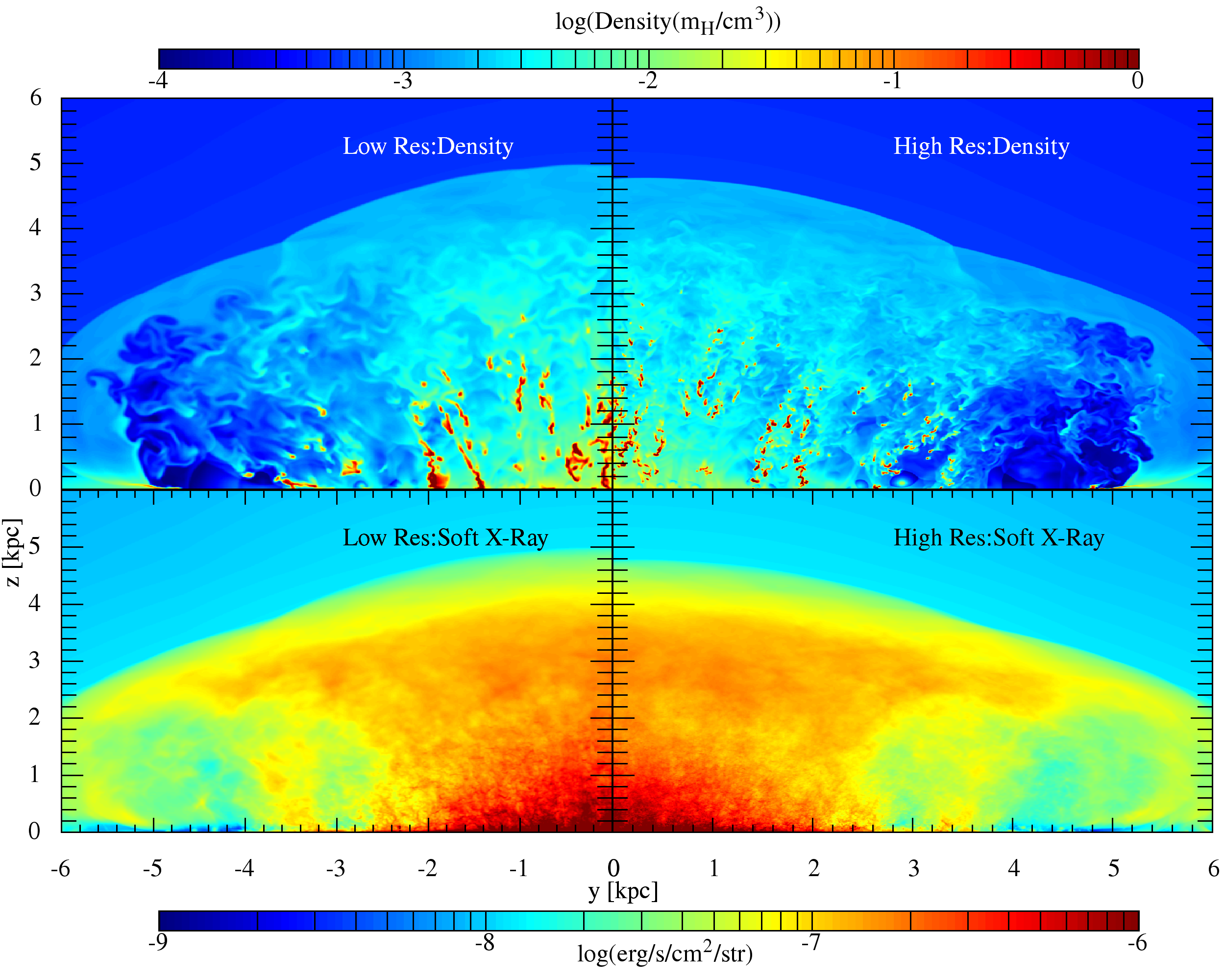}
 \caption{Density in the $x=0$ plane(top) and soft X-Ray ($0.5-2.0$ keV) surface brightness along $x=0$ line of sight(bottom) at $5$ Myr. The left panels depict the low resolution case($12$ pc) which has been studied elsewhere in the paper, while the right ones indicate the high resolution run($6$ pc) meant for comparison. The plots seem qualitatively similar. It should be noted that density contours in the HR case have smaller clouds.} 
\label{fig:densLRHR}
 \end{figure*}

We conducted resolution study for $L42$, DI case. Using same injection points as for $L42$ we ran the set-up for a period of $10$ Myr but with a resolution of $6$ pc (referred to as HR). Since the energy and mass deposition rates are identical to $L42$ we expect broad agreement with our conclusions  \textit{viz} formation of clumps of disc gas and filamentary nature of the soft X-Ray surface brightness. This is borne by density and soft X-Ray surface brightness contours(\ref{fig:densLRHR}). Further scrutiny of the density plots reveals that HR produces smaller clouds than $L42$. This, however, is a known effect and is a result of higher cooling which also results in a slower shock speed for HR.

\begin{figure*}
\includegraphics[width=7cm]{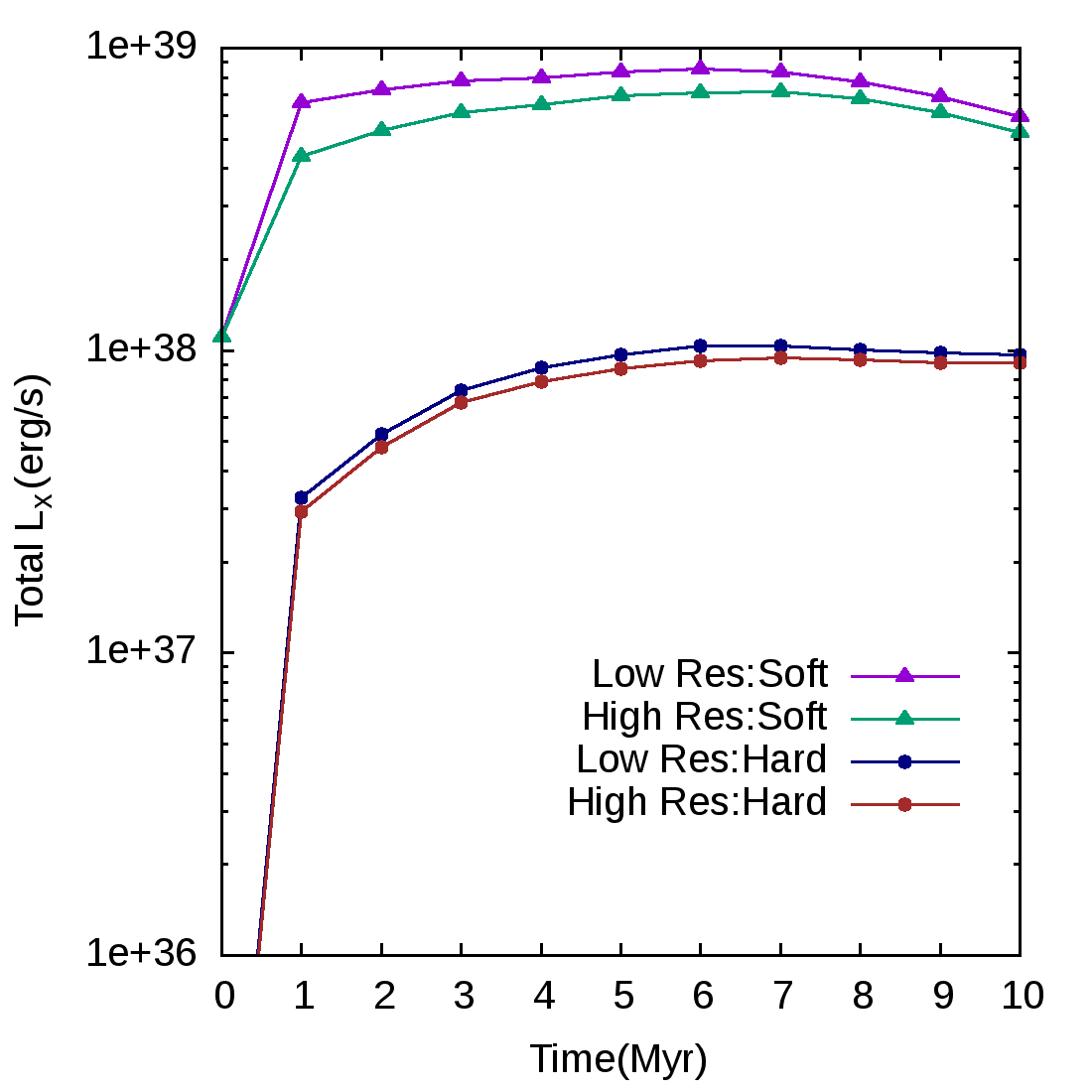}
 \caption{The total soft X-Ray emission is higher in case of LR by a factor $\leq$ 1.2 as compared to HR.} 
\label{fig:totLxLRHR}
 \end{figure*}

Since the clouds are smaller in HR we expect a relatively smaller total soft X-Ray emission. The soft X-ray luminosity is about $1.2$ times higher for $L42$ than HR(\ref{fig:totLxLRHR}), whereas the hard X-ray luminosity remains roughly the same.s

%%%%%%%%%%%%%%%%%%%%%%%%%%%%%%%%%%%%%%%%%%%%%%%%%%

% Don't change these lines
\bsp	% typesetting comment
\label{lastpage}
\end{document}